\documentclass[letterpaper,11pt]{article}
\pdfoutput=1 

\usepackage{jheppub} 
\usepackage[utf8]{inputenc}
\usepackage[T1]{fontenc} 
\usepackage[dvipsnames]{xcolor}
\usepackage{multirow}
\usepackage{slashed}
\usepackage[normalem]{ulem}
\usepackage{sidecap}
\usepackage{mathtools}
\usepackage{caption}
\usepackage{subcaption}
\usepackage[toc]{appendix}

\usepackage{amsmath}
\usepackage{graphicx}
\usepackage{fancyhdr}
\usepackage{multirow}
\usepackage{booktabs}
\usepackage{makecell}
\usepackage{tabulary}
\usepackage{feynmp}
\title{A model for mixed warm and hot right-handed neutrino dark matter}

\author[a]{Ma\'ira Dutra}
\author[b]{, Vin\'icius Oliveira}
\author[b]{, C. A de S. Pires}
\author[c,d]{, Farinaldo S. Queiroz}

\affiliation[a]{Ottawa-Carleton Institute for Physics,
Carleton University, 1125 Colonel By Drive, Ottawa, Ontario K1S 5B6, Canada}
\affiliation[b]{Departamento de F\'isica, Universidade Federal da Para\'iba, Caixa Postal 5008, 58051-970,
Jo\~ao Pessoa, PB, Brazil}
\affiliation[c]{International Institute of Physics, Universidade Federal do Rio Grande do Norte,
Campus Universit\'ario, Lagoa Nova, Natal-RN 59078-970, Brazil}
\affiliation[d]{Departamento de F\'isica, Universidade Federal do Rio Grande do Norte, 59078-970, Natal, RN, Brasil}

\emailAdd{mdutra@physics.carleton.ca}
\emailAdd{vlbo@academico.ufpb.br}
\emailAdd{cpires@fisica.ufpb.br}
\emailAdd{farinaldo.queiroz@iip.ufrn.br}

\usepackage{calligra}
\usepackage{calrsfs}
\DeclareMathAlphabet{\mathcalligra}{T1}{calligra}{m}{n}
\DeclareMathAlphabet{\pazocal}{OMS}{zplm}{m}{n}

\def\to{\rightarrow}

\def\figureautorefname~#1\null{Fig.\,#1\null}
\def\tableautorefname~#1\null{Tab.\,#1\null}

\def\equationautorefname~#1\null{Eq.\,(#1)\null}

\abstract{
We discuss a model where a mixed warm and hot keV neutrino dark matter rises naturally. We arrange active and sterile neutrinos in the same $SU(3)_L$ multiplet, with the lightest sterile neutrino being dark matter. The other two heavy sterile neutrinos, through their out-of-equilibrium decay, contribute both to the dilution of dark matter density and its population, after freeze-out. We show that this model features all ingredients to overcome the overproduction of keV neutrino dark matter, and explore the phenomenological implications for Big Bang Nucleosynthesis and the number of relativistic degrees of freedom.
}

\begin{document}

\today
\maketitle
\flushbottom

\section{Introduction}
\label{sec:intro}

Although there is no doubt about the existence of dark matter (DM)\cite{Hinshaw:2012aka,Aghanim:2018eyx}, we have no idea about its nature. There are compelling pieces of evidence that dark matter may be composed of elementary particles, all based on its gravitational interaction with ordinary matter. Such particles must be electrically neutral (at least effectively) and cosmologically stable. Dark matter is also crucial for evolution of structure formation as we observe today. In general, dark matter candidates are classified as hot dark matter (HDM), warm dark matter (WDM), or cold dark matter (CDM) depending on their free-streaming around the period of structure formation. Structure formation requirements do not allow for dark matter to be comprised of mostly HDM \cite{Primack2001}. 

Weakly interacting massive particles (WIMPs) are by far the most extensively studied class of CDM as the correct dark matter abundance is easily reproduced with cross sections around the weak scale \cite{Bertone:2004pz}. They have been extensively searched for by many experiments (direct and indirect detection, and colliders) with no success \cite{Arcadi:2017kky}. The null results reported thus far motivate us to explore alternative candidates. As CDM faces problems at small-scale astrophysical scales, mixed populations of dark matter are well motivated \cite{Bull:2015stt}. Right-handed neutrinos (from now on sterile neutrinos) with mass around keV scale are suitable candidates.  Sterile neutrinos arise in many popular extensions of the standard model (SM) such as left-right model \cite{Mohapatra:1974gc,Senjanovic:1975rk,Senjanovic:1978ev}, B-L model  \cite{Davidson:1978pm,Mohapatra:1980qe,Appelquist:2002mw,Basso:2008iv,Khalil:2010iu} and in  a particular version of gauge models based in the $SU(3)_C \times SU(3)_L \times U(1)_N$ symmetry \cite{Singer:1980sw,Montero:1992jk,Foot:1994ym}.

Although interesting alternatives to CDM candidates, keV sterile neutrinos are usually overproduced in simplified models \cite{Adhikari:2016bei}. In this case, the most plausible way to \textit{dilute} this dark matter population and obtain the correct abundance is through entropy injection \cite{Kolb:1990vq}. 
In general, entropy can be injected into the early universe when 
a long-lived particle, a \textit{diluton}, that decouples while relativistic, dominates the energy density of the universe and decays once non-relativistic \cite{Kolb:1990vq}. Thus, a successful keV dark matter model should feature a long-lived particle that plays the role of such diluton. The natural diluton candidates are the right-handed neutrino themselves. The lightest right-handed neutrino is stable, while the other two act as the diluton \cite{Dror:2020jzy, Bezrukov:2009th, Nemevsek:2012cd}. This apparent easy solution faces solid constraints from Big Bang Nucleosynthesis (BBN) and the Cosmic Microwave Background (CMB). 

The study of keV neutrino dark matter has been discussed elsewhere in simplified models, our goal here is to embed this mechanism in a UV complete model, which is well motivated for other theoretical reasons as family replication\cite{Foot:1992rh} and electric charge quantization\cite{deSousaPires:1998jc,deSousaPires:1999ca}. We will discuss this keV neutrino dark matter in a model based on the $SU(3)_C \times SU(3)_L \times U(1)_N$ gauge symmetry, 331 for short. There are several ways to arrange the fermion generations in this gauge symmetry. The different ways give rise to different models. Here we will focus on the  331$\nu_R$ \cite{Singer:1980sw,Montero:1992jk,Foot:1994ym}, which features right-handed neutrinos in the same $SU(3)_L$ multiplet where resides the active neutrinos \cite{Dias:2005yh}.

In this work, we check under which conditions the 331$\nu_R$ accommodates a successful keV dark matter candidate. To do so, we invoke a discrete symmetry to guarantee the stability of the lightest sterile neutrino. We calculate the relativistic freeze-out of the lightest sterile neutrino and consider the heavier ones as dilutons. As we are dealing with an extended gauge sector there are new interactions that impact the sterile neutrino abundance, differing from past keV neutrino dark matter studies. We highlight that the two heavy sterile neutrinos dilute and produce dark matter. To test our model, we estimate 
the free-streaming length of our dark matter candidate and find that we have a mixed WDM+HDM scenario, which successfully obeys the constraints stemming from CMB and BBN. 

This work is organized as follows. In \autoref{sec:model} we present the key aspects of the model; in \autoref{sec:relic} and \autoref{sec:coupled} we address the production mechanisms; in \autoref{sec:constraints} we outline the viable parameter space; lastly in we draw our conclusions in \autoref{sec:conclusions}.

\section{The essence of the 331$\nu_R$ model}
\label{sec:model}

\subsection{Particle content}
In the 331$\nu_R$ model the lepton generations are arranged as,
\begin{eqnarray}
L_{l} = \left (
\begin{array}{c}
\nu_{l_L} \\
e_{l_L} \\
(\nu_{l_R})^{c}
\end{array}
\right )\sim(1\,,\,3\,,\,-1/3)\,,\,\,\,e_{lR}\,\sim(1,1,-1),
 \end{eqnarray}
where $l = e,\,\mu,\, \tau$ refers to the three generations.

In the quark sector, anomaly cancellation requires  the first two generations of  quarks to  come in the anti-triplet while the third generation in
 triplet representation as,
\begin{eqnarray}
&&Q_{iL} = \left (
\begin{array}{c}
d_{i} \\
-u_{i} \\
d^{\prime}_{i}
\end{array}
\right )_L\sim(3\,,\,\bar{3}\,,\,0)\,\,\,,u_{iR}\,\sim(3,1,2/3),\,\,\,\nonumber \\
&&\,\,d_{iR}\,\sim(3,1,-1/3)\,,\,\,\,\, d^{\prime}_{iR}\,\sim(3,1,-1/3),\nonumber \\
&&Q_{3L} = \left (
\begin{array}{c}
u_{3} \\
d_{3} \\
u^{\prime}_{3}
\end{array}
\right )_L\sim(3\,,\,3\,,\,1/3)\,,\,\,\,u_{3R}\,\sim(3,1,2/3),\nonumber \\
&&\,\,d_{3R}\,\sim(3,1,-1/3)\,,\,u^{\prime}_{3R}\,\sim(3,1,2/3)
\label{quarks} 
\end{eqnarray}
where $i=1,2$. The new quarks ($u^{\prime}\,\,\,,\,\, d^{\prime}$)
have the usual electric charges.

The gauge sector of the model is formed by the standard gauge bosons, $A$, $W^{\pm}$ and $Z$,  and  five others called  $W^{\prime \pm}$, $U^0$, $U^{0 \dagger}$ and $Z^{\prime}$. The interactions of these gauge bosons with matter can be found in Ref. \cite{Hoang:1995vq}.

The scalar sector of the original version involves three  scalar triplets, namely $\eta = (\eta ^0\,,\, \eta^- \,,\, \eta^{0 \prime})^T\sim ({\bf 1}\,,\,{\bf 3}\,,\,-1/3)$, $\chi=(\chi^0 \,,\, \chi^- \,,\,\chi^{0 \prime})^T \sim ({\bf 1}\,,\,{\bf 3}\,,\,-1/3)$, and $\rho=(\rho^+ \,,\,\rho^0 \,,\,\rho^{+ \prime})^T \sim ({\bf 1}\,,\,{\bf 3}\,,\,2/3) $. The quantum numbers under $SU(3)_C \times SU(3)_L \times U(1)_N$ are displayed in parenthesis

With such scalar content, when the symmetry $SU(3)_C \times SU(3)_L \times U(1)_N$ is spontaneously broken to $SU(3)_C \times U(1)_{em}$, all particles acquire masses at tree level, except the neutrinos.

We adopt the following vacuum structure,
\begin{eqnarray}
\langle \eta \rangle_0 = \left (
\begin{array}{c}
 \frac{v_\eta}{\sqrt{2}} \\
 0\\
 0 
\end{array}
\right ),\,\langle \rho \rangle_0 =\left (
\begin{array}{c}
0 \\
\frac{v_\rho}{\sqrt{2}} \\
0
\end{array}
\right ) ,\, \langle \chi \rangle_0 = \left (
\begin{array}{c}
0 \\
0 \\
\frac{v_{\chi^{\prime}}}{\sqrt{2}}
\end{array}
\right )\,, \label{VEVstructure} 
\end{eqnarray}
which implies that $v^2_\eta +v^2_\rho=(246)^2$GeV$^2$, to make sure that $M_{W}^2 = g^2 (v_\eta^2 +v_\rho^2)/4$ in agreement with the Standard Model. 

To further simplify the  model, we assume the following discrete symmetry transformation over the full Lagrangian,
\begin{eqnarray}
&&\left( \,\eta,\,\rho,e_{lR},\,u_{aR},\,d_{aR}\right) \rightarrow-\left( \,\eta,\,\rho,e_{lR},\, u_{aR},\,d_{aR}\right),
\nonumber \\
\label{discretesymmetryI}
\end{eqnarray}
where $a=1,2,3$. This discrete symmetry ($Z_2$) will play an important role in the dark sector of the model, as we shall see. With such matter and scalar content we build the following  Yukawa interactions invariant under the gauge symmetry,
\begin{eqnarray}
{\mathcal L}^Y &=&\lambda^1_{ij}\bar Q_{iL}\chi^* d^{\prime}_{jR}+\lambda^2_{33}\bar Q_{3L}\chi u^{\prime}_{3R}+ \lambda^3_{ia}\bar Q_{iL}\eta^* d_{aR}+\nonumber \\
&&\lambda^4_{3a}\bar Q_{3L}\eta u_{aR}+ \lambda^5_{ia}\bar Q_{iL}\rho^* u_{aR}+\lambda^6_{3a}\bar Q_{3L}\rho d_{aR}+\nonumber \\
&& G_{l l^{\prime}}\bar L_{l_L} \rho e_{l^{\prime}_R}+\mbox{H.c},
\label{yukawa}
\end{eqnarray}
which generate masses for all fermions, with the exception of neutrinos.

We remark that while the SM does not contain any dark matter candidate, the 331$\nu_R$ model poses three candidates, namely, $\eta^{0 \prime}$, $U^0$ or $\nu_R$, which are mutually exclusive. $U^0$ is underabundant, while $\eta^{0 \prime}$ and $\nu_R$ are viable multi-TeV dark matter candidates \cite{deS.Pires:2007gi,Mizukoshi:2010ky,Alvares:2012qv,Profumo:2013sca,Kelso:2013nwa,Dong:2014wsa,Cogollo:2014jia,Kelso:2014qka,Alves:2016fqe,Dong:2017zxo}. Concerning $U^0$, despite of being an interesting candidate, unfortunately it does not provide the correct abundance (it is under-abundant), while  $\nu_R$ did not receive any attention until now.   In other words, this is the first time that right-handed  neutrino is being treated as dark matter in the 331$\nu_R$.

Before considering $\nu_R$ as dark matter candidate we discuss in the next section how to generate  masses for the neutrinos in the model. 
\subsection{ Neutrino Masses}
Right-handed neutrinos are hypothetical particles and their masses are free parameters that may take
a wide range of values varying from eV up to GUT scale. In the original version of the 331$\nu_R$ model neither $\nu_L$ nor $\nu_R$ gain masses. The most immediate way of providing masses for them is through effective dimension-5 operators\cite{Dias:2005yh}. In the case of left-handed neutrinos, this operator is constructed with the scalar triplet $\eta$ and the lepton triplet $L$,
\begin{eqnarray}
{\mathcal L}_{\nu_L}&&=	\frac{f_{l l^{\prime}}}{\Lambda}\left( \overline{L^C_l} \eta^* \right)\left( \eta^{\dagger}L_{l^{\prime}} \right)+\mbox{H.c}.
	\label{5dL}
\end{eqnarray}	
According to this operator, when $\eta^0$ develops a VEV, $v_\eta$, the left-handed neutrinos develop  Majorana mass terms,
\begin{eqnarray}
(m_{\nu_L})_{l l^{\prime}}=\frac{f_{l l^{\prime}}v^2_\eta}{\Lambda}.
	\label{nuL}
\end{eqnarray}

Regarding right-handed neutrinos,  the dimension-5  operator that give them mass is constructed with the scalar triplet $\chi$ and the lepton triplet $L$,
\begin{eqnarray}
{\mathcal L}_{\nu_R}&&=	\frac{h_{l l^{\prime}}}{\Lambda}\left( \overline{L_{l}^C} \chi^* \right)\left( \chi^{\dagger}L_{l^{\prime}} \right)+\mbox{H.c}.
	\label{5dR}
\end{eqnarray}	
When $\chi^{\prime 0}$ develops a VEV, $v_{\chi^{\prime}}$, this effective operator  provides  Majorana masses for the right-handed neutrinos,
\begin{eqnarray}
(m_{\nu_R})_{l l^{\prime}}=\frac{h_{l l^{\prime}}v^2_{\chi^{\prime}}}{\Lambda}.
	\label{nuR}
\end{eqnarray}  
Once $v_{\chi^{\prime}} > v_\eta$, then $m_{\nu_R} > m_{\nu_L}$. Thus, light right-handed neutrinos is a natural result of the model. 

Observe that the discrete symmetry discussed above avoids the operator $\sim \frac{1}{\Lambda}( L \eta)( \chi L)$ which would generate  mixing among active and sterile neutrinos. This is a particularly interesting result, as it renders the lightest right-handed neutrino automatically stable and suitable to be our dark matter candidate.  The ways these operators can be realized are not relevant for the issue we are going to address here. However, for the sake of completeness, by invoking the existence of a sextet of scalar with mass belonging to the GUT scale, we can realize such operators according to type II seesaw mechanism \cite{Ma:1998dn}, as done in Refs. \cite{Montero:2001ts,Cogollo:2009yi,Dong:2008sw}.

\subsection{Main interactions}
To work with the physical neutrinos we have to diagonalize  $m_{\nu_L}$ and $m_{\nu_R}$. From now on we refer to the active  physical neutrinos as $\nu_L =(\nu_1\,,\,\nu_2\,,\, \nu_3)_L^T$, and to the physical sterile ones as  $N_L=(N^C_1\,,\, N^C_2 \,,\, N^C_3)_L^T$. To simplify even more we assume $\nu_R$ in a diagonal basis. 

We now present  the interactions involving sterile neutrinos that matter for us here:

\begin{equation} \label{eq:1}
\mathcal{L}_{W'}  =  - \frac{g}{\sqrt{2}} \overline{N}_L \gamma^{\mu} l_L W^{'+}_{\mu} + \text{H.c.}
\end{equation}

\begin{equation} \label{eq:2}
\mathcal{L}_{U^0}  =  -\frac{g}{\sqrt{2}} \overline{\nu}_L \left( U^T_{PMNS} \right) \gamma^{\mu} N_L U^{0}_{\mu} +\text{H.c.}
\end{equation}

\begin{eqnarray} \label{eq:3}
\mathcal{L}_{Z'} && =  - \frac{g}{2C_W} \left( \frac{\left( 1 - 2 S^2_W \right)}{\sqrt{3- 4S_W^2}} \left[\overline{\nu}_L  \gamma^{\mu} \nu_L \right] -   \frac{2 C_W^2}{\sqrt{3- 4S_W^2}} \left[\overline{N}_L  \gamma^{\mu} N_L  \right] \right) Z^{\prime}_{\mu} \nonumber \\ 
&&- \frac{g}{4C_W \sqrt{3-4 S^2_W}} \bar l \gamma^\mu \left( (3-4C^2_W) + \gamma_5  \right)l Z^{\prime}_{\mu},
\end{eqnarray}
where $C_W=\cos(\theta_W)$ and $S_W=\sin(\theta_W)$ with $\theta_W$ being the Weinberg angle and $ l=( e \,,\, \mu \,,\, \tau )^T$.

The dominant interactions involving quarks that matter for the calculation of the dark matter abundance are
\begin{eqnarray} \label{eq:4}
\mathcal{L}_{Z'} && =  - \frac{g}{2C_W} \frac{\sqrt{ 3 - 4 S^2_W} }{3} \left[\overline{u}_{L}  \gamma^{\mu} u_{L} \right]Z^{\prime}_\mu -  \frac{g}{ 2 C_W}\frac{2(1- S_W^2)}{\sqrt{3- 4S_W^2}} \left[\overline{t}_{L}  \gamma^{\mu} t_{L}  \right] Z^{\prime}_{\mu} \nonumber \\ 
 &&   - \frac{g}{2C_W} \frac{\sqrt{3 - 4 S^2_W}}{3} \left[\overline{d}_{L}  \gamma^{\mu} d_{L} \right]Z^{\prime}_\mu -  \frac{g}{ 2 C_W}\frac{2(1- S_W^2)}{\sqrt{3- 4S_W^2}} \left[\overline{b}_{L}  \gamma^{\mu} b_{L}  \right] Z^{\prime}_{\mu} ,
\end{eqnarray}
where $u=(u\,,\,c)^T$ and $d=(d\,,\,s)^T$.

\section{Relic abundance of a light sterile neutrino}
\label{sec:relic}

\begin{figure}[!t]
    \centering
    \includegraphics[scale=0.3]{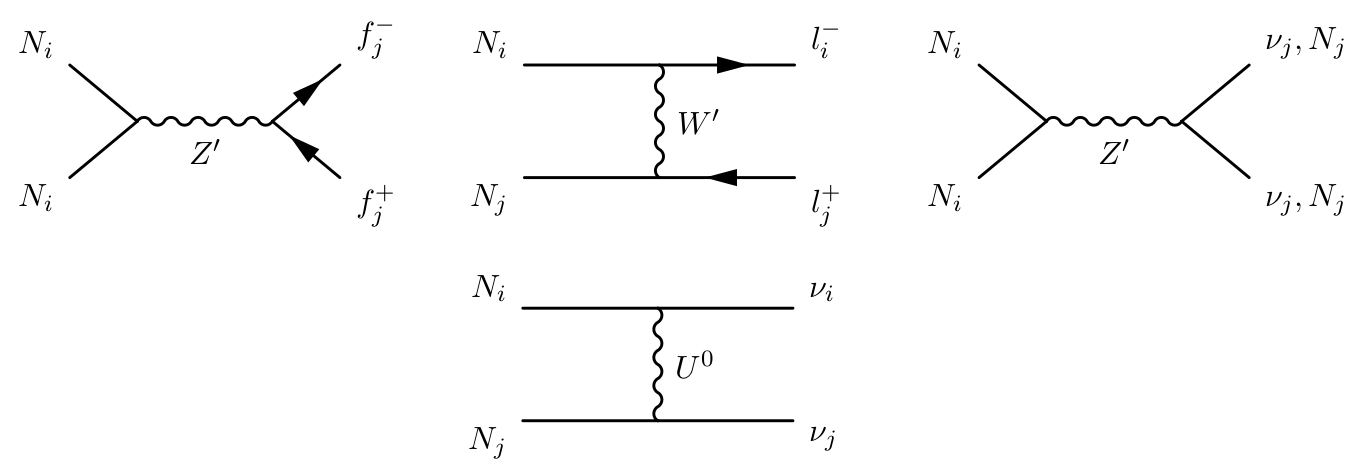}
\caption{Processes that contribute to $N_i$ freeze-out where $f$ represents the charged fermions of standard models and $l$ represents the charged leptons with $i,j = 1, 2, 3$.}
\label{Fig: Process}
\end{figure}

We assume that $N_1$ is the lightest of the sterile neutrinos, being in principle a
dark matter candidate. We therefore
check under which conditions it thermalizes with species of the standard model bath in the early universe and calculate its relic abundance accordingly. 

In the context of the $331\nu_R$ model, the sterile neutrinos are able to thermalize with the standard fermions through exchanges of $W^{\prime}$, $Z^{\prime}$ and $U^0$, as depicted in \autoref{Fig: Process}. This happens whenever their interaction rates $\Gamma_{N_i}(T)$ are faster than the Hubble rate $H(T)$ at a given temperature $T$:
\begin{eqnarray}
    \frac{\Gamma_{N_i}\left( T \right)}{H \left( T \right)} \gg 1\,.
\end{eqnarray}

The rate at which $N_i$ self-annihilate into species $3$ and $4$, with masses $m_{N_3}$ and $m_{N_4}$, is given by $\Gamma_{N_i} (T) = n_{N_i}^{eq}(T) \langle \sigma v \rangle $, with $n_{N_i}^{eq}(T)$ their equilibrium number density and a thermally averaged annihilation cross-section given by
\begin{equation}\label{thermally_avg}
    \langle \sigma v \rangle \equiv \frac{1}{(n_{N_i}^{eq}(T))^2} \frac{\cal{S}}{32 \left(2 \pi \right)^6} T \int ds \frac{\sqrt{\lambda \left(s, m_{N_1}^2,m_{N_1}^2\right)}}{s} \frac{\sqrt{\lambda \left(s, m_{N_3}^2,m_{N_4}^2\right)}}{\sqrt{s}} K_1 \left(\frac{\sqrt{s}}{T} \right) \int d \Omega |\mathcal{M}|^2 \,,
\end{equation}
where $\cal{S}$ is the symmetrization factor, $s$ is the Mandelstam variable, $\lambda \left( x, y, z \right)$ is the Källen function, $K_i$ is the modified Bessel function of the second kind of order $i$, $\Omega$ is the solid angle between initial and final states in the center of mass frame, and $|\mathcal{M}|^2$ the (not averaged) squared amplitude of the process.

As usual, we compute the freeze-out temperatures $T_f$ at which the sterile neutrinos decouple from the thermal bath by equaling $n_{N_i}^{eq}(T_f) \langle \sigma v \rangle (T_f) = H(T_f)$ for the main processes. For simplicity but without loss of generality for our purposes, we will assume the hierarchy $m_{N_3} \gg m_{N_2} \gg m_{N_1}$. As a consequence, we can neglect co-annihilation processes \cite{Griest:1990kh}.

\begin{figure}[t!]
    \centering
    \includegraphics[scale=0.3]{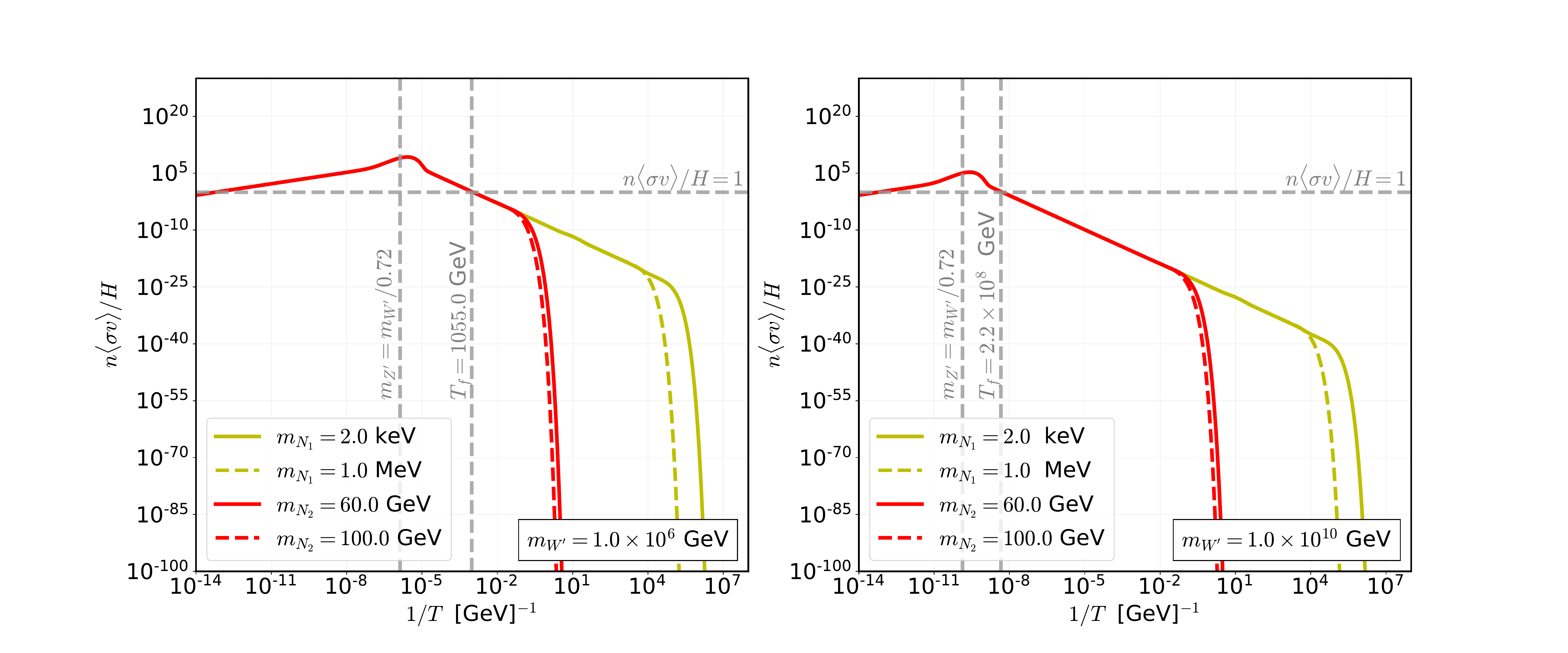}
    \caption{Ratio between the interaction rates $\Gamma \equiv n \langle \sigma v \rangle$ and the Hubble rate $H$ for $N_1$ (yellow) and $N_2$ (red) annihilations as a function of the inverse of temperature. The left panel accounts for $m_{W^\prime} = 1.0 \times 10^6$ GeV and the right panel, for $m_{W^\prime} = 1.0 \times 10^{10}$ GeV. The dashed horizontal line represents $\Gamma/H = 1$, below which the sterile neutrinos are decoupled. The dashed vertical lines indicates where $T=m_{Z^\prime}$ and $T=T_f$.}
    \label{fig:thermalize}
\end{figure}

In \autoref{fig:thermalize} we show the ratio between the annihilation rates of $N_1$ (in yellow) and of $N_2$ (in red) and the Hubble rate in the standard case of a radiation-dominated era as a function of the inverse of temperature\footnote{For our numerical results, we have used the {\tt CUBA} library~\cite{Hahn:2004fe}.} for $m_{W^\prime} = 1.0 \times 10^6 \ GeV$ (left panel) and $m_{W^\prime} = 1.0 \times 10^{10} \ GeV$ (right panel). To obtain the curves of \autoref{fig:thermalize} we consider all annihilation processes shown in \autoref{Fig: Process}. This analysis is also applicable to the case of $N_3$, but for the purpose of this paper we do not need to take it into account, as we will see later. As shown in \autoref{fig:thermalize}, the freeze-out temperature $T_f$ increases with the mass of the mediator. In our model, the gauge bosons have similar masses, and from now on we  assume $m_{W^\prime}=m_{U^0} \simeq 0.72 \times m_{Z^{\prime}}$\cite{Hoang:1995vq}. We can see in \autoref{fig:thermalize} that the maximum of the ratio happens when the gauge bosons are produced on-shell, making the decoupling to take place close to that scale.

We have chosen in \autoref{fig:thermalize} masses of $N_1,$ $N_2$, and $W^\prime$ close to the values which will provide the right amount of relic abundance for $N_1$, as we will show later. In such a case, we can see that $N_1$ and $N_2$ freeze-out almost at the same time (when $n_{N_i}\langle \sigma v \rangle \sim H$, as indicated by the dashed horizontal line), with $T_f \simeq 1.0 \times 10^3$ GeV for $m_{W^\prime} = 1.0 \times 10^6 \ GeV$ (left panel) and $T_f \simeq 2.2 \times 10^8$ GeV for $m_{W^\prime} = 1.0 \times 10^{10} \ GeV$ (right panel). Thus, both $N_1$ and $N_2$ decouple while still relativistic. As shown in \autoref{fig:thermalize} the freeze-out temperature $T_f$ strongly depends of mediator mass, for that benchmark value of mass. When the masses of $N_1$ and $N_2$ increase (indicated by the yellow and red dashed curves in \autoref{fig:thermalize}, respectively) the Boltzmann suppression (which occur when $m_{N_i} \sim T$) occurs earlier, as shown in \autoref{fig:thermalize}. Given that in our model the couplings between sterile neutrinos, gauge bosons, and standard fields are all of order $\mathcal{O} (10^{-1})$, the only way of avoiding thermalization of sterile neutrinos is by invoking gauge bosons heavier than about $\mathcal{O} (10^{16})$ GeV.

We remark that among all processes contributing to the freeze-out of $N_1$ and $N_2$, the most relevant are annihilations into SM charged leptons and active neutrinos. In the relevant limit of $m_{N_i} \ll T \ll m_{W^\prime}$, they happen at the following rate:
\begin{equation}\label{thermally_aprx}
    n_{N_i}\langle \sigma v \rangle \ \simeq \ \frac{16}{9} \frac{T^5 G_F^2}{\zeta \left(3 \right) (2 \pi)^4} \ \left( \frac{m_W}{m_{W^{\prime}}}\right)^4 \left( 64 -  8 \left( \frac{m_{W^\prime}}{m_{Z^\prime}} \right)^2 \right) \,.
\end{equation}

We have found the typical freeze-out temperature to be given by
\begin{equation}\label{free_out_temperature}
T_f \simeq 1.7 \left( \frac{m_{W^\prime}}{m_W} \right)^{4/3} g_{e}^{1/6} \left( T_f \right) \text{MeV} \,,
\end{equation}
where $g_{e}(T_f)$ is the number of energetic degrees of freedom at freeze-out, with $g_e(T_f)\sim 100$ for $T_f>100$ GeV. We have checked that our numerical solution shown in \autoref{fig:thermalize} is in a good agreement with the estimation above.

We can therefore conclude that $N_1$ would be a cosmic relic which was once thermalized, and proceed with the computation of their final abundance in order to determine whether it can constitute the cosmological dark matter.

\subsection{Relativistic freeze-out} 

The final relic abundance of $N_1$ is defined to be
\begin{equation}
    \frac{\Omega_{N_1}^0 h^2}{0.12} \simeq \frac{m_{N_1}}{1 \text{GeV}} \frac{Y_{N_1}^0}{4.34 \times 10^{-10}}\,,
\end{equation}
where the label "$^0$" indicates quantities as measured today, with $\Omega^0_{DM} h^2 \simeq 0.12$ being inferred by the Planck satellite \cite{Aghanim:2018eyx}, and $Y_i \equiv n_i/s$ is the yield of a species $i$, with $s$ the entropy density in a comoving volume.

In this work, we are interested in the case of a light sterile neutrino dark matter. For $N_1$ thermally produced the lower limit on 
its mass is $m_{N_1} \gtrsim 2.0$ keV \cite{Seljak:2006qw}. As we have just seen, for the mass hierarchy $m_{N_1}, m_{N_2} \ll m_{W^\prime}$, both $N_1$ and $N_2$ decouple from the thermal bath almost simultaneously and at high temperatures ($T \sim m_{W^\prime}/100$). The yield of a Majorana neutrino that decouples while ultra-relativistic is given by 
\begin{equation}\label{Eq:YrelFO}
    Y_{N_i}(T) = \frac{135 \zeta(3)}{4\pi^4 g_s(T_f)}\,,
\end{equation}
for temperatures $T$ below their freeze-out temperature $T_f$, where $g_s(T_f)$ is the number of entropic degrees of freedom at freeze-out. 

Therefore, if $Y_{N_1}^0 = Y_{N_1}(T_f)$, the agreement with the relic abundance constraint,
\begin{equation}\label{Eq:DilutedRelic}
    \frac{\Omega_{N_1}^0 h^2}{0.12} \simeq \left(\frac{m_{N_1}}{1 \text{keV}}\right) \left(\frac{1000}{g_s(T_f)}\right)\,,
\end{equation}
would require an unreasonable amount of relativistic degrees of freedom by the time of freeze-out in order for a light sterile neutrino to not overclose the universe.

To the best of our knowledge, the only way of depleting the yield of a decoupled species is by considering entropy production after freeze-out. As it is well known \cite{PhysRevD.31.681}, a long-lived particle that decoupled while ultra-relativistic can dominate the cosmic expansion before decaying, thus injecting a sizable amount of entropy into the thermal bath. It is therefore interesting to notice that the heavier sterile neutrinos are natural candidates to deplete the yield of $N_1$. For simplicity, we will investigate the out-of-equilibrium decay of $N_2$ as the source of entropy production, while assuming that $N_3$ is heavy enough as to not affect our analysis.

It is straightforward to see that an increase of total entropy $S$ in a comoving volume after the freeze-out of any relic, by a factor of $\Delta \sim S (T_0)/S (T_f)$, will dilute its yield by the same factor: $Y_{N_i} (T_0) = Y_{N_i} (T_f)/\Delta$. For freeze-out happening at TeV scale, our model provides $g_s(T_f) \sim 100$. Thus, $N_1$ with mass in the keV-MeV range requires an entropy injection of $\Delta = 10 - 100$ in order to be a viable dark matter candidate. As we show in what follows, though, the out-of-equilibrium decay of $N_2$ contributes in a non-trivial way to the final abundance of $N_1$ in the context of the $331\nu_R$ model.

\subsection{Non-thermal production}

As we have just discussed, the out-of-equilibrium decay of $N_2$ into species of the thermal bath dilutes the abundance of our dark matter candidate, $N_1$. However, in the $331\nu_R$ model, tree-body decays of $N_2$ into $N_1$ can be sizable, which could potentially repopulate (and overclose) the universe with dark matter. Therefore, the final relic abundance of $N_1$ will have a (thermal) contribution from the relativistic freeze-out and also a (non-thermal) contribution from the tree-body decays of $N_2$ into $N_1$:
\begin{equation}
\Omega_{N_1}^0 h^2 = \Omega_{N_1}^0 h^2\Big|_{thermal} + \Omega_{N_1}^0 h^2\Big|_{non-thermal}\,.
\end{equation}

Let us parametrize the total decay width of $N_2$ in terms of the partial width into $N_1$, $\Gamma_{N_2}^{(N_1)}$:
\begin{equation}\label{eq:alpha}
\Gamma_{N_2} \equiv (1+\alpha)\Gamma_{N_2}^{(N_1)} \,.
\end{equation}

The dimensionless parameter $\alpha = \sum_i \Gamma_{N_2}^{i}/\Gamma_{N_2}^{(N_1)}$ contains all other channels which do not involve $N_1$ as final product \footnote{ All these channels are mediated
by the new gauge bosons and scalars of the model. They involve only charged leptons and hadrons as final products, such as for instance $N_2 \rightarrow l^{\pm} +\text{hadrons}$. Given the complexity of our scalar sector, these processes might be abundant and dominate the decay of $N_2$, thus making the model-dependent parameter $\alpha$ sizable.}. Of course, $N_2$ must decay into species which thermalize with the SM bath in order to dilute $N_1$ . Decay channels into $331\nu_R$ states do not necessarily thermalize, but here we will treat $\alpha$ as a free parameter encoding only decay channels which instantaneously thermalize with the SM bath.

We have found that the leading contributions to $\Gamma_{N_2}^{(N_1)}$ are the three-body decays into charged leptons and neutrinos. In the limit $m_{N_1}, m_l \ll m_{N_2} \ll m_{W^\prime}$, we have
\begin{equation}\label{Eq:Decays}
\Gamma_{N_2}^{(N_1)} \equiv \Gamma_{N_2 \to \mu e N_1} + \Gamma_{N_2 \to \nu_\mu \nu_e N_1} \approx \frac{G_F^2}{96\pi^3} m_{N_2}^5 \left(\frac{m_W^4}{m_{W^\prime}^4} + \frac{m_W^4}{m_U^4}\right)\,.
\end{equation}

In the next section, we develop the tools needed to properly dealing with the competing effects of the $N_2$ out-of-equilibrium decays.

\section{Coupled evolution of sterile neutrinos}\label{sec:coupled}

Let us now discuss how to properly find the final relic abundance of our dark matter candidate, the keV scale sterile neutrino $N_1$. The relativistic freeze-out of both $N_1$ and $N_2$, as well as the non-thermal production and dilution of $N_1$ due to the out-of-equilibrium decay of $N_2$, can be taken into account by solving the following coupled Boltzmann fluid equations for the yields of $N_1$ and $N_2$ \footnote{We recall that we consider the hierarchy $m_{N_1} \ll m_{N_2} \ll m_{N_3}$, such that $N_3$ decays at much higher temperatures and does not significantly affect the lighter sterile neutrinos.}:
\begin{equation}\label{Y_equation}
\begin{split}
    \frac{d Y_{N_1}}{da} & = \frac{R_{N_1}( a, Y_{N_1}, Y_{N_2})}{s \ H(a) \ a} - \frac{Y_{N_1}}{S} \frac{dS}{da} \\ 
    \frac{d Y_{N_2}}{da} & =  \frac{R_{N_2}( a, Y_{N_1}, Y_{N_2})}{s \ H(a) \ a} - \frac{Y_{N_2}}{S} \frac{dS}{da} \,,
\end{split}
\end{equation}
where the scale factor $a$ is used as a time parameter.

The relativistic freeze-out and the non-thermal production of $N_1$ are accounted for by the first term in the right hand side of the equation above, whereas the second term accounts for the dilution of $Y_{N_1}$ after the entropy production.

The reaction rate densities $R_{N_1,N_2}$ contain all processes that can change the number of $N_1$ and $N_2$ in a comoving volume. We have found the following leading contributions: 
\begin{equation}\label{rate_reaction}
\begin{split}
    R_{N_1} & \ \approx  \  - s^2  \langle \sigma v \rangle_{N_1 N_1} \left( Y_{N_1}^2 - \left(Y_{N_1}^{(eq)}\right)^2 \right) + s \ \langle \Gamma_{N_2}^{(N_1)} \rangle \left(Y_{N_2} - Y_{N_1} \frac{Y_{N_2}^{(eq)}}{Y_{N_1}^{(eq)}} \right)\\
    R_{N_2} & \ \approx \  - s^2  \langle \sigma v \rangle_{N_2 N_2} \left( Y_{N_2}^2 - \left(Y_{N_2}^{(eq)}\right)^2 \right) - s \ \langle \Gamma_{N_2}^{(N_1)} \rangle \left(Y_{N_2} - Y_{N_1} \frac{Y_{N_2}^{(eq)}}{Y_{N_1}^{(eq)}} \right)  \\ 
    & \hspace{.8cm} - \alpha \ s \ \langle \Gamma_{N_2}^{(N_1)} \rangle \left(Y_{N_2} - Y_{N_2}^{(eq)} \right)\,.
\end{split}
\end{equation}

The terms proportional to $\langle \sigma v \rangle_{N_i N_i}$ represent the annihilations into SM leptons (see \autoref{thermally_aprx}) and their backreactions. We can represent $\langle \sigma v \rangle_{N_i N_i}$ as the sum of the channels that contribute for it:
\begin{eqnarray}
    \langle \sigma v \rangle_{N_i N_i} = \langle \sigma v \rangle_{N_i N_i \to e \bar{e}} + \langle \sigma v \rangle_{N_i N_i \to \mu \bar{\mu}} + \langle \sigma v \rangle_{N_i N_i \to \tau \bar{\tau}} +  \langle \sigma v \rangle_{N_i N_i \to \nu_e \bar{\nu_e}} +\\ \nonumber \langle \sigma v \rangle_{N_i N_i \to \nu_\mu \bar{\nu_\mu}} + \langle \sigma v \rangle_{N_i N_i \to \nu_\tau \bar{\nu_\tau}}\,.
\end{eqnarray}

The other terms represent the contribution of the $N_2$ decays and inverse decays. We recall that $\alpha$ encodes the channels without $N_1$ as a decay product, assuming that they all thermalize instantaneously. We therefore see that the decay of $N_2$ into $N_1$ couples their evolution in the early universe, even if the entropy production were negligible.

In order to inject a significant amount of entropy into the thermal bath after decaying, $N_2$ must be significantly long-lived and have dominated the total energy density of the universe. This is indeed a natural consequence of our framework, since $N_2$ decouples while ultra-relativistic -- which means that $\rho_{N_2}/\rho_R \propto m_{N_2} a$ once it becomes non-relativistic, with $\rho_R$ the energy density of radiation. The Hubble rate in \autoref{Y_equation} will be therefore given by
\begin{equation}\label{Hubble_matter}
    H (a) = \frac{ \sqrt{\rho_R(a) + \rho_{N_2}(a) }}{\sqrt{3} M_{Pl}}\,,
\end{equation}
where $M_{Pl} \simeq 2.4 \times 10^{18}$ GeV is the reduced Planck mass.

From the temperature at which $N_2$ starts dominating the energy density, $T_i$, until its complete decay, at the so-defined reheat temperature $T_{RH}$, the universe would have therefore undergone an early matter-dominated era. The duration of such an era is determined by the amount of entropy produced, $T_i/T_{RH} \propto \Delta$, and the reheat temperature is found to be given by \cite{Cosme:2020mck}
\begin{equation}\label{reheating_temperature}
    T_{RH} = \left( \frac{5 \pi^2}{72} g_e \left( T_{RH} \right) \right)^{-1/4} \sqrt{\Gamma_{N_2} M_{Pl}}\,.
\end{equation}

In order to not jeopardize the BBN predictions \cite{Hasegawa:2019jsa}, we must ensure $T_{RH}\gtrsim 4$ MeV. This guarantees that  $N_2$ decays before the weak decoupling of active neutrinos and all the standard leptons thermalize.

The rate of injection of entropy due to the decay of $N_2$ is given by \cite{PhysRevD.31.681},
\begin{equation}\label{Entropy_Equation}
\frac{dS}{da} \ = \ f_T \ \frac{\Gamma_{N_2}}{H} \ \frac{ \ \rho_{N_2} \ a\left( t \right)^2}{T} \,,
\end{equation}
where $f_T$ represents the fraction of the decay products of $N_2$ that thermalize in the plasma \cite{Patwardhan:2015kga,Fuller:2011qy}. The fraction $f_{NT}$ of decay products that do not thermalize will populate  the sea of decoupled relativistic species (contributing to $\Delta N_{eff}\neq 0$, see \autoref{Sec:Neff}).  

It is therefore convenient to rewrite the energy density of $N_2$ in terms of $f_T$ and $f_{NT}$:
\begin{equation}
\begin{split}
\rho_{N_2} = & \overbrace{\left[ f_T^l \cdot Br( N_2 \to \mu e N_1) + f_T^\nu \cdot Br( N_2 \to \nu_\mu \nu_e N_1) + Br(N_2 \to others) \right]}^{f_T} \cdot  \rho_{N_2} +\\ 
& \overbrace{\left[ f_{NT}^l \cdot Br( N_2 \to \mu e N_1) + f_{NT}^\nu \cdot Br( N_2 \to \nu_\mu \nu_e N_1) \right]}^{f_{NT}} \cdot  \rho_{N_2}\,,
\end{split}
\end{equation}
where $Br(N_2 \to others) = 1/(1+1/\alpha)$ (see \autoref{eq:alpha}) is the branching ratio into all decay channels without $N_1$ in final states.

Since $T_{RH} < T_f$, the $N_1$ produced via decay will not be able to thermalize anymore. This is why such dark matter population is said to be non-thermal. On the other hand, above $4$ MeV, all SM leptons are able to thermalize. It is then easy to see that $f_T^l = f_T^\nu = 2/3$, whereas $f_{NT}^l = f_{NT}^\nu = 1/3$, so that 
\begin{equation}\label{fraction}
\begin{split}
f_T &= \frac{\alpha+2/3}{1+\alpha}\\ 
f_{NT} &= \frac{1/3}{1+\alpha}\,.
\end{split}
\end{equation}

Finally, since $\rho_{N_2}$ and $\rho_R$ evolve non-trivially during the evolution of $N_1$ and $N_2$, the set of \autoref{Y_equation} must be solved together with the following Boltzmann fluid equations: 
\begin{equation}\label{Energy_equation}
\begin{split}
& \frac{d \rho_{N_2}}{dt} + 3 H \rho_{N_2} = - \rho_{N_2} \Gamma_{N_2} \\
& \frac{d \rho_{R}}{dt} + 4 H \rho_{R} =  \rho_{N_2} \Gamma_{N_2} \,.
\end{split}
\end{equation}

\subsection{Numerical results}\label{numerical}

We numerically solve the set of equations (\ref{Y_equation}), (\ref{Hubble_matter}), (\ref{Entropy_Equation}), and (\ref{Energy_equation}). For numerical convenience, we re-scale the scale factor by $A \equiv  a T_{RH}$, the energy density of $N_2$ by $\Phi_{N_2} \equiv \rho_{N_2} a^4$, and the energy density of radiation by $\Phi_R \equiv \rho_R a^4$.

Regarding the initial conditions, at $A=A^I \ll 1$ (the actual value does not change results), the yields of $N_1$ and $N_2$ follow their equilibrium values. The freeze-out of the sterile neutrinos take place while they are still relativistic, during the radiation era. From $A^I$ until the moment when $N_2$ becomes non-relativistic, at $A = A_{NR}$, there is no entropy production and we just need to solve the coupled set of \autoref{Y_equation} without the last terms. In this case the Hubble rate is the usual one, $H \propto T^2/M_{Pl}$.

For $A \geqslant A_{NR}$, we consider the full set of equations. We follow Ref. \cite{Cosme:2020mck} and assume an inflationary model that yield an inflationary reheat temperature of $T_{RH}^{Inf} \simeq 7 \times 10^{15}$ GeV, which implies $S^I = S^{NR} \simeq 897$ and $\Phi_R^I = \Phi_R^{NR} \simeq 1790$. At $A_{NR}$, both $Y_{N_1}$ and $Y_{N_2}$ are given by \autoref{Eq:YrelFO}. Since at this point $N_2$ is non-relativistic, we have
\begin{equation}
\Phi_{N_2}^{NR} = m_{N_2} Y_{N_2}(T_f) \frac{S^{NR} A^{NR}}{T_{RH}}\,.
\end{equation}

\begin{figure}[t!]
    \centering
  \includegraphics[scale = 0.36]{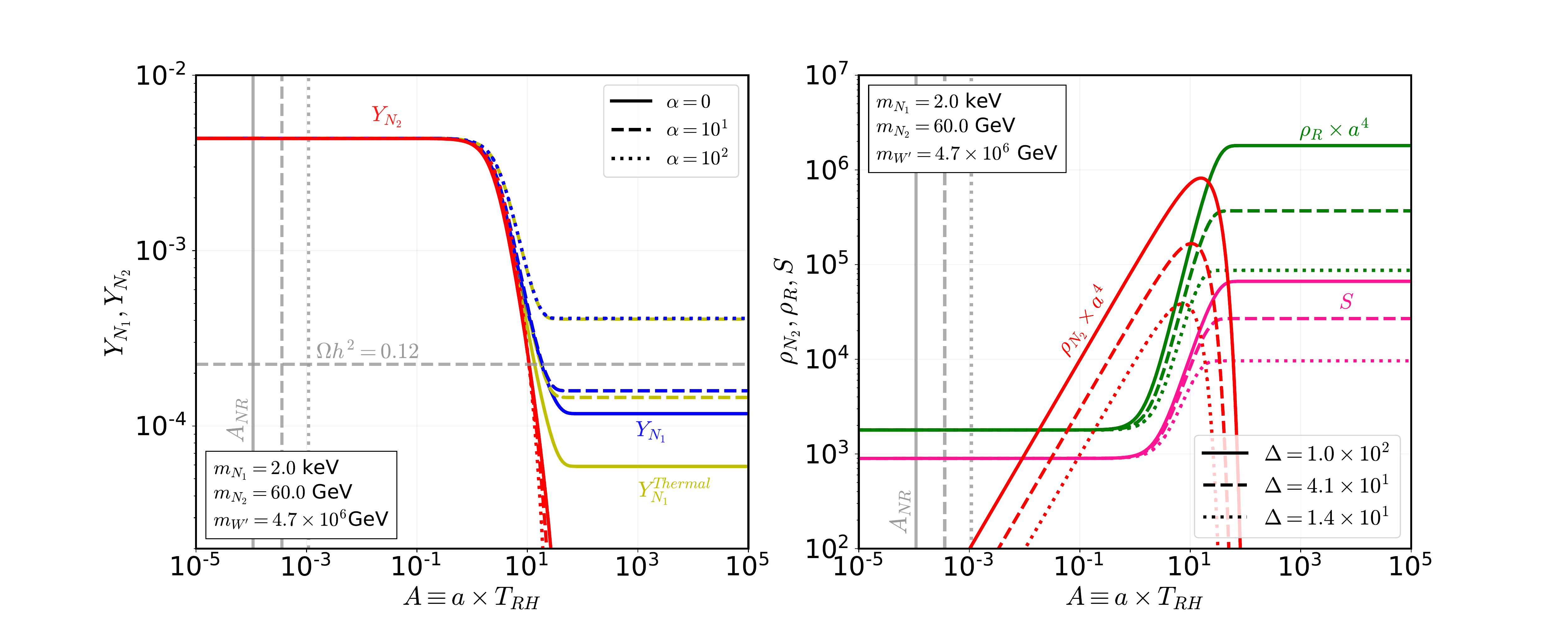} 
    \caption{Evolution of coupled equations (\ref{Y_equation}), (\ref{Entropy_Equation}), and (\ref{Energy_equation}) for $\alpha = 0, 10, 100$
    (solid, dashed and dotted curves,    respectively). Left panel: Evolution of total (thermal and non-thermal contribution) $Y_{N_1}$ (blue),  only thermal contribution $Y_{N_1}^{Thermal}$ (yellow) and $Y_{N_2}$ (red).
    Right panel: Curves of entropy $S$ (pink), $\rho_{N_2} \times a^4$ (red) and $\rho_{R} \times a^4$ (green), also shown are how the value of $\Delta$ changes as we increase $\alpha$ ($\Delta = 100, 41$ and $14$, for respective $\alpha$).}
    \label{coupled_equations}
\end{figure}

In \autoref{coupled_equations} we present the evolution of the set of equations (\ref{Y_equation}), (\ref{Entropy_Equation}), and (\ref{Energy_equation}) for $\alpha = 0, 10$ and $100$ (continuous, dashed and dotted curves, respectively). On the left panel of \autoref{coupled_equations} we show the full solutions for $Y_{N_1}$ (blue curve) and $Y_{N_2}$ (red curve), as well as the solution for $Y_{N_1}$ in the absence of non-thermal contribution, $Y_{N_1}^{Thermal}$ (yellow curve). On the right panel we show the solutions
for entropy $S$ (pink curve) and for the quantities $\rho_{N_2}a^4$ (red curve) and $\rho_{R}a^4$ (green curve) and we can observe that when $N_2$ becomes non-relativistic (at $A_{NR}$, as shown) the energy density of radiation is still greater than the energy density of $N_2$. However the ratio between the energy density of non-relativistic $N_2$ and the energy density of radiation evolves like $\rho_{N_2}/ \rho_R \propto am_{N_2}$, then $\rho_{N_2}$ can dominate the energy density of the universe if $N_2$ is sufficiently
long lived, as shown in \autoref{coupled_equations}. The complete decay of $N_2$ correspond in \autoref{coupled_equations} to the abrupt decrease of $Y_{N_2}$ on the left panel (or $\rho_{N_2}a^4$ on the right panel).

We can observe in the right panel of \autoref{coupled_equations} that when $N_2$ decays completely, the entropy is increased by a factor $\Delta$ and, as expected, it keeps constant before and after the decaying of $N_2$. The increase of entropy dilutes the abundance of $N_1$, as we have discussed, and we can observe this behavior from the evolution of the $Y_{N_1}$ (blue curves) on the left panel in the \autoref{coupled_equations}. When $N_2$ decays completely, the injection of entropy ceases and $Y_{N_1}$, $S$ and $\rho_Ra^4$ levels off. 

As shown in \autoref{coupled_equations} the free parameter $\alpha$ plays an important role in the dilution of $N_1$. Naively, one would expect that the dilution would increase with $\alpha$, since more thermalized decay channels are allowed. However, since $\Gamma_{N_2}$ increases with $\alpha$, increasing $\alpha$ makes $N_2$ to decay earlier, such that it does not dominate the evolution of the universe long enough for a significant entropy injection to take place.
We can observe this behavior on the right panel of \autoref{coupled_equations}: when $\alpha = 0$ (solid
curves), $\rho_{N_2} > \rho_R$ for a much longer period
and inject more entropy ($\Delta = 100$) than when 
we take $\alpha = 10^2$ (dotted curves) and obtain $\Delta = 14$. As the injection of entropy (parameterized by $\Delta$) is responsible for diluting
$N_1$, this explains why the final value of $Y_{N_1}$ is higher as we increase $\alpha$.

The free parameter $\alpha$ has also an important role on the non-thermal contribution to the abundance of $N_1$. As shown in the left panel of \autoref{coupled_equations} there is a gap between $Y_{N_1}$ (thermal and non-thermal contribution) and $Y_{N_1}^{Thermal}$ (only thermal contribution) which means that the $N_1$ produced via $N_2$ decays is responsible for increasing
the abundance of $N_1$ from $Y_{N_1}^{Thermal}$ to $Y_{N_1}$. We can observe that this gap decreases
as we increase $\alpha$. This behavior is explained by the fact that when $\alpha$ increases, it allows $N_2$ to decay in another particles beyond $N_1$. According to our computation, for $\alpha = 10^2$ we have that the non-thermal contribution is approximately $1\%$, while for $\alpha = 0$ the non-thermal contribution is approximately $50\%$.

\section{Viable parameter space}
\label{sec:constraints}

\begin{figure}[t!]
    \centering
  \includegraphics[scale=0.6]{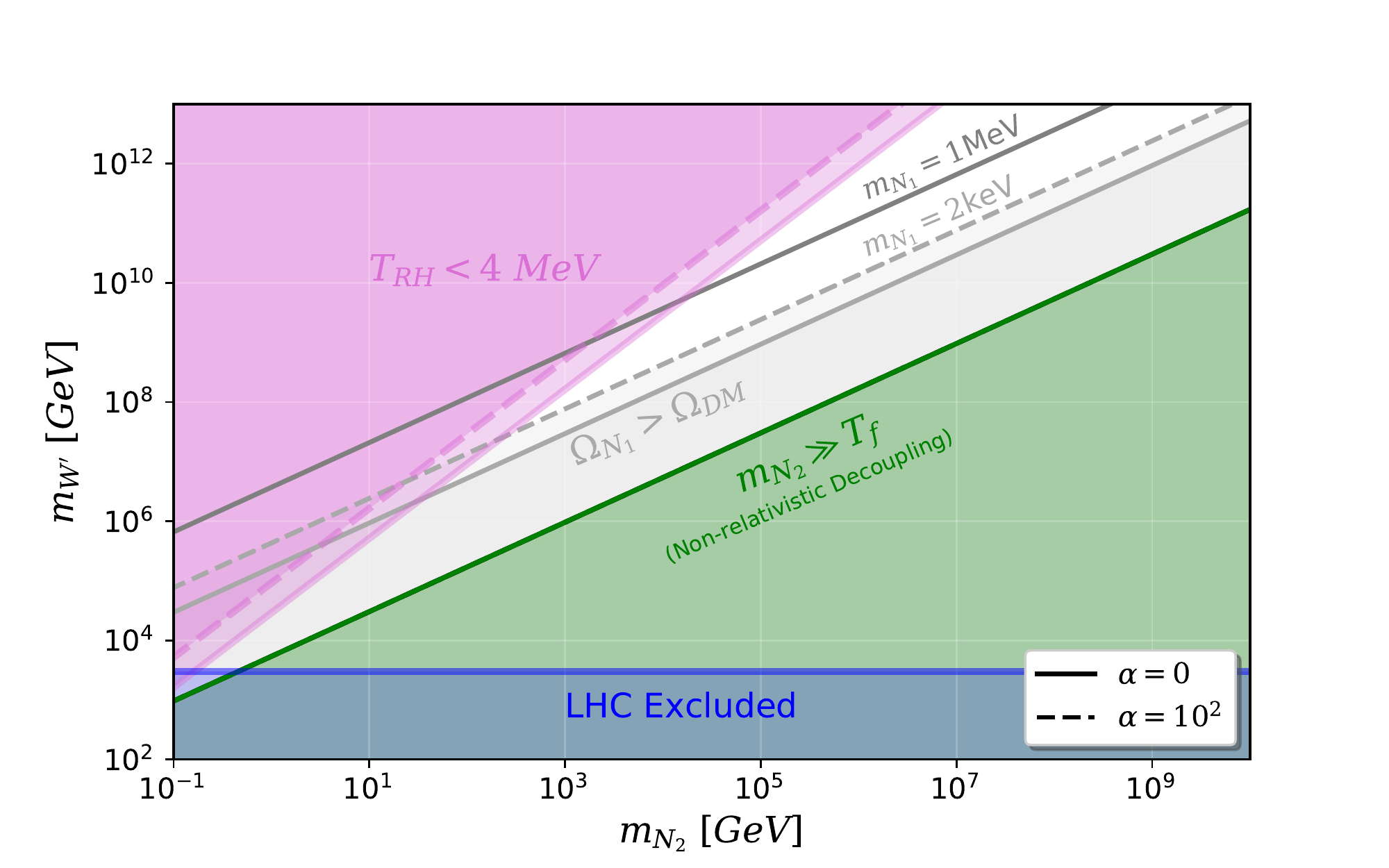} 
    \caption{Viable parameter space in the plane $(m_{W^\prime}, m_{N_2})$. The correct dark matter relic density is achieved along the light-grey ($m_{N_1} = 2$keV ) and grey lines ($m_{N_1} = 1$MeV). The continuous lines are for $\alpha =0$ and the dashed lines, for $\alpha = 10^2$. The pink and blue regions are excluded respectively by BBN ($T_{RH} < 4 MeV$) and LHC ($m_{W^\prime} < 3\ GeV$). In the green region, the necessary requirement of ultra-relativistic decoupling of $N_2$ does not hold.}
    \label{bound_mass}
\end{figure}

In this section we obtain the constraints on the parameter space $(m_{W^{\prime}}\,,\,m_{N_2})$, that makes $N_1$ produced by means of relativistic freeze-out and diluted from $N_2$ decay a realistic DM candidate. In \autoref{bound_mass} we present these constraints for $\alpha = 0$ (solid curves)
and $\alpha = 100$ (dashed curves).

Our first constraint is on the reheating temperature, given in \autoref{reheating_temperature}, which is the temperature soon after the decay of $N_2$. As we have pointed out, $N_2$ must decay prior to the active neutrinos decoupling as to not disturb the BBN predictions, such that we need to ensure $T_{RH} \gtrsim 4$ MeV.
In \autoref{bound_mass} the region in pink is excluded because it gives $T_{RH} < 4$ MeV. As  $T_{RH} \propto \sqrt{\Gamma_{N_2}}$ and $\Gamma_{N_2}$ increases with $\alpha$, this bound, translated to an upper limit on $M_{W^\prime}$, is weakened as we increase $\alpha$.

As we discussed above, $N_2$ must decouple ultra-relativistic in order to produce enough entropy and sufficiently dilute $N_1$. This criterion rule out the green-shaded region in \autoref{bound_mass}, in which $m_{N_1}\gg T_f$ with 
$T_f$ given by \autoref{free_out_temperature}. This
bound is independent of the parameter $\alpha$. Finally, the
LHC constraint over the mediator mass, $ m_{W^\prime} \gtrsim 3 \ TeV $ \cite{Cao:2016uur,Coutinho:2013lta,Cogollo:2020afo,deMelo:2021ers}, is indicated by the blue-shaded region.  

The correct relic abundance of $N_1$ today, $\Omega_{N_1} h^2 \simeq 0.12$, as constrained by the Planck satellite \cite{Aghanim:2018eyx}, is ensured by demanding
\begin{equation}\label{condition_relic_abundance}
    Y_{N_1}^0 \simeq 4.25 \times 10^{-4} \left( \frac{1 \ \text{keV}}{m_{N_1}}\right)\,.
\end{equation}

In order to take into account the thermal and non-thermal contributions, we find $Y_{N_1}^0$
by solving  \autoref{Y_equation} numerically.   For this  we developed a Python algorithm which
searches the values of $m_{N_2}$ and $m_{W^\prime}$ obeying \autoref{condition_relic_abundance}. The contours of correct relic density are shown for $m_{N_1} = 2$keV (light-grey curve) and $m_{N_1} = 1$MeV (grey curve), for $\alpha = 0$ (continuous) and $\alpha = 100$ (dashed). The region below the light-grey (grey) contours are excluded by Planck as they overclose the universe. As the purpose of this paper is to study $N_1$ as a viable DM candidate, we are interested in the light-grey (grey) curves themselves. As we can see, the heavier the dark matter, the heavier the mediators need to be for the agreement with Planck.

\subsection{Contribution to $\Delta N_{eff}$}
\label{Sec:Neff}

The amount of relativistic particles at matter-radiation equality epoch contributes to the number of relativistic degrees of freedom and directly affects the CMB power spectrum. An important parameter in this regard
is the effective number of neutrino species defined as  $ N_{eff}=\frac{8}{7}(\frac{11}{4})^{4/3}(\frac{\rho_{rad}-\rho_\gamma}{\rho_\gamma})$, where $\rho_{rad}$ and $\rho_\gamma$ are respectively the total radiation and photon energy density.
The current value for $N_{eff}$ from Planck 2018+BAO \cite{Aghanim:2018eyx} is $N_{eff} = 2.99 \pm 0.17$ which is in agreement with the standard model prediction $N_{eff}^{SM} = 3.0440$ \cite{Froustey:2020mcq}. The change in the amount of radiation is quantified by means of the  parameter $\Delta N_{eff} \equiv N_{eff} - N_{eff}^{SM}$.

Since the fraction of $N_1$ produced non-thermally becomes non-relativistic at temperatures of the order $\mathcal{O}\left(eV\right)$ 
(see Appendix \ref{Ap:1}), it increases the value of $N_{eff}$. 
We remark that the $N_1$ produced thermally and then diluted does not contribute to $\Delta N_{eff}$ because it becomes non-relativistic before equality (see Appendix \ref{Ap:1}). 

The contribution to
$\Delta N_{eff}$ in our model comes therefore from the non-thermal population of $N_1$ and is mainly controlled by the parameter $\alpha$:
\begin{align}\label{effective_neutrino}
    \Delta N_{eff} & = \frac{4}{7} \left( \frac{11}{4} \right)^{4/3} \frac{1}{3} \left( Br \left( N_2 \to \mu e N_1\right) + Br \left( N_2 \to \nu_\mu \nu_e N_1 \right) \right) \left( \frac{g_{s, equality}^4}{g_{s, RH}} \right)^{1/3}
    \\ & = \frac{4}{7} \left( \frac{11}{4} \right)^{4/3} \frac{1}{3} \left( 1 + \alpha 
    \right)^{-1} \left( \frac{g_{s, equality}^4}{g_{s, RH}} \right)^{1/3}\,. \nonumber
\end{align}

According to this, $\alpha = 0 $ yields $\Delta N_{eff} \simeq 2.05$ which is excluded by the bounds discussed above. In other words, we need $\alpha > 0$. Future experiments as CMB-S4 will have a sensitivity to constraint $\Delta N_{eff} = 0.060$ (at $95 \% $ C.L.) \cite{Abazajian:2019eic}. In the absence of evidence for $\Delta N_{eff} \neq 0$, we will have the limit $\Delta N_{eff} < 0.06$ which  requires $\alpha \gtrsim 33$.

\subsection{Structure formation (free-streaming)}

As dark matter can be classified as cold (CDM), warm (WDM) or hot (HDM) according to its free-streaming $\lambda_{fs}$ \cite{Merle:2013wta}, in this section we will roughly estimate the free-streaming of $N_1$.

The free-streaming at the epoch of matter-radiation equality is an important parameter that allows us to
understand how the first structures were formed. It is given by
\cite{kolb_early_1990}

\begin{equation}
    \lambda_{fs} \equiv \int_{t_{prod}}^{t_e} \frac{ \langle v(t) \rangle}{a(t)} dt \,,
\end{equation}
where $t_{prod}$ is the production time, $t_e$ 
the equality time, and $\langle v(t) \rangle$ 
the mean velocity of $N_1$. 

The $N_1$ population thermally produced and then diluted due to an injection of entropy is "colder" than what it would be if only thermally produced. The relation between the temperature of the decoupled $N_1$ and that of the plasma is affected
by the injection of entropy, parametrized by $\Delta$, and is given by
\begin{equation}\label{cooling}
 \frac{T_{N_1}}{T} = \left( \frac{1}{\Delta}\frac{g_s}{g_{s,b}}\right)^{1/3}\,,
\end{equation}
where the subscript "$b$" refers to any epoch prior to the dilution. 

The thermally produced fraction of a keV $N_1$, which is  relativistic at production, has its free-streaming suppressed to the scale of $\mathcal{O} \left(0.1 \text{Mpc} \right)$ due the dilution $\Delta$, being therefore classified as warm or even cold dark matter (see for instance Refs. \cite{Patwardhan:2015kga,Merle:2013wta}).

On the other hand,  the free-streaming of $N_1$ produced non-thermally is given by  \cite{Borgani:1996ag, Lin:2000qq}
\begin{align}
\lambda_{fs} \ & \simeq \int_{0}^{t_{e}} \frac{v(t)}{a(t)} dt\\ \nonumber & \simeq 2 v_0 t_{e} \left( 1 + z_{e}\right)^2  ln \left( \frac{1}{v_0 \left( 1+ z_{e} \right)} + \sqrt{1 +\frac{1}{v_0^2 \left( 1 + z_{e} \right)^2}} \right) \,,
\end{align} 
where $v_0$ is the initial velocity of $N_1$ and $z_{e}$ is the red-shift at equality. 

Since the non-thermal population of $N_1$ is produced at $T_{RH}$ through 3-body decay of $N_2$, and in the limit of massless final states, we have $p_{prod}^{N_1} \simeq m_{N_2}/3$. For temperatures below $T_{RH}$, the universe is 
radiation-dominated and the temperature red-shifts as $a = a_0 \times T_0/T$. We have therefore
\begin{equation}
    v_0 \equiv \frac{p_{prod}^{N_1}}{m_{N_1}} a_{prod} \ \simeq \ \frac{m_{N_2}}{3 m_{N_1}} \frac{T_0}{T_{RH}}\,.
\end{equation}

For $m_{N_1} = 2.0$ keV, $m_{N_2} = 60.0$ GeV and $m_{W^\prime} = 4.7 \times 10^6$ GeV we obtain $\lambda_{fs} \simeq 2.4 Mpc$. The non-thermal fraction of $N_1$ is therefore classified as HDM. 

Our non-thermal component can compose up to $50\%$ of the total abundance (for $\alpha = 0$) and $1\%$ (for $\alpha = 10^2$). As the CMB constraint requires $\alpha>0$, we have a scenario of mixed warm-hot DM. Such a scenario has important implications for structure formation and can address some small-scale problems such as the core-cusp \cite{10.1111/j.1365-2966.2011.20200.x}. Warm dark matter candidates can delay the formation of structures, with lighter dark matter increasing the delay. The recent measurement of the 21-cm absortion signal due to the light of the first stars, reported by the EDGES Collaboration \cite{Bowman:2018yin}, can provide lower bounds on the mass of non-cold dark matter candidates. If $N_1$ is mainly warm, this observation implies roughly $m_{N_1}\lesssim 3$ keV \cite{Chatterjee:2019jts}. A rigorous approach to the implications of mixed DM in the structure formation may also put bounds on the parameter $\alpha$ \cite{Dayal:2015vca,Schneider:2018xba}. However this requires the computation of the fluctuation of the power spectrum for our scenario and is beyond the scope of this paper.

\section{Conclusions}
\label{sec:conclusions}

We investigated the possibility of having a viable mixed warm and hot keV neutrino dark matter in a model based on the $SU(3)_L$ gauge group. Active and sterile neutrinos are arranged in the same $SU(3)_L$ multiplet, with the lightest sterile neutrino being dark matter. Its abundance is set by interactions with Standard Model particles controlled by the new gauge bosons rising from the extended gauge sector. Its  stability is warranted by a discrete symmetry that prevents mixing between active and sterile neutrinos, making the model safe from otherwise stringent bounds, such as X-rays.

We have shown that the sterile neutrinos easily thermalize with the Standard Model bath via exchanges of these heavy new gauge bosons, unless their masses are at the GUT scale. When the sterile neutrinos are much lighter than the gauge bosons, they decouple nearly at temperatures much larger than their masses, rendering them ultra-relativistic, and therefore overproduced, at the thermal freeze-out.

In this work, we have shown that this common issue with keV right-handed neutrino is solved within this $SU(3)_L$ gauge group. Since the heavier sterile neutrinos decouple while ultra-relativistic, and are long-lived due to the heavy gauge bosons mediating their decays, they eventually dominate the cosmic expansion after freeze-out. Their out-of-equilibrium decay into SM bath species and also into keV right handed neutrino dark matter. Hence they contribute both to the dilution of warm dark matter population and as well as to its non-thermal HDM population.

Our scenario of mixed warm-hot dark matter is amenable cosmological constraints. As this entropy injection episode should take place before BBN, constraints are derived on the masses of the heavy sterile neutrino states, and gauge bosons. We found that they must be larger than the TeV scale. Moreover, the hot dark matter can also increase $N_{eff}$ in a detectable way, and for this reason we imposed  $\alpha \gtrsim 33$ to make sure that the heavy sterile neutrino decay mostly into particles of the $SU(3)_L$ gauge group other than dark matter.

In summary, we conclude that our model can  successfully host a mixed warm plus hot dark matter setup in agreement with existing bounds.

\acknowledgments 

 M.D. acknowledges the support of the Arthur B. McDonald Canadian Astroparticle Physics Research Institute and of the Natural Sciences and Engineering Research Council of Canada. C.A.S.P  is supported by the CNPq research grants No. 304423/2017-3. V.O acknowledges CNPq for financial support.  FSQ is supported by the S\~ao Paulo Research Foundation (FAPESP) through grant 2015/158971, ICTP-SAIFR
FAPESP grant 2016/01343-7, CNPq grants 303817/2018-6 and 421952/2018 - 0, and the Serrapilheira Institute (grant
number Serra - 1912 - 31613).

\appendix
\section{$N_1$ temperature at the matter-radiation equality}\label{Ap:1}

Here we will check when $N_1$ produced thermally and non-thermally becomes non-relativistic and if they contribute to $\Delta N_{eff}$ at CMB.

Following \cite{Borgani:1996ag, Hasenkamp:2012ii}, the momentum of $N_1$ produced non-thermally $ p_{N_1}^{NT}$ by $N_2$ decay gets red-shifted 

\begin{equation}
    p_{N_1}^{NT}\left( T \right) \ = \ |p_{N_1}^{NT}\left( T_D \right)| \frac{a_{D}}{a\left(T \right)}\,, 
\end{equation}
where $a_D$ is the scale factor at the moment of decay and $p_{N_1}^{NT}\left( T_D \right)$ is the momentum of non-thermal $N_1$ when it is produced.
Assuming that $m_{N_2} \gg m_{N_1}, m_{e,\mu, \nu_e, \nu_{\mu}}$, which imply in

\begin{equation}\label{Nontherm.nonrelat}
     |p_{N_1}^{NT}\left( T_D \right)| \ \simeq \ \frac{m_{N_2}}{3}
\end{equation}
then,
\begin{equation}\label{momentum_non_thermally}
     p_{N_1}^{NT}\left( T \right) \ = \ \frac{m_{N_2}}{3} \frac{a_{D}}{a\left(T \right)} \,.
\end{equation}

We assume that $N_1$ produced via decay becomes non-relativistic at some temperature $T_{nr}$ (with the respective scale factor $a_{nr}$). This happen when $p_{N_1}^{NT} \left( T_{nr} \right) = m_{N_1}$ \cite{Hasenkamp:2012ii}. Then, from \autoref{momentum_non_thermally} we have

\begin{equation}\label{nr_momentum_condition}
     m_{N_1} \ = \ \frac{m_{N_2}}{3} \frac{a_{D}}{a_{nr}}\,.
\end{equation}

We can represent the red-shift when non-thermal $N_1$ becomes non-relativistic as
\begin{equation}\label{nr_redshift}
Z_{nr}^{NT} + 1 \equiv \frac{a_0}{a_{nr}} =  \frac{a_0}{a_D} \frac{a_D}{a_{nr}}
\end{equation}
For a universe dominated by radiation the temperature red-shifting as $T = T_0 \times a_0/a$ and using \autoref{nr_momentum_condition} we obtain
\begin{equation} \label{Z_non-thermally}
   Z_{nr}^{NT} + 1 = \frac{3 m_{N_1}}{m_{N_2}} \frac{T_D}{T_0} \left( \frac{g_s (T_D)}{g_s (T_0)} \right)^{1/3}\,,
\end{equation}
where,
\begin{equation}\label{Temperature_decouple}
    T_D \ \simeq T_{RH} = \ 116 MeV \ \left( 1 + \alpha \right)^{1/2} \ \left(\frac{70}{g_e \left( T_{RH}\right)} \right)^{1/4}\left(\frac{m_{N_2}}{100GeV} \right)^{5/2} \left(\frac{10^6 GeV}{M_{W^\prime}} \right)^2 \,.
\end{equation}
As a result of \autoref{Z_non-thermally} we obtain that $N_1$ produced non-thermally becomes non-relativistic at temperatures $\mathcal{O} \left( eV \right)$ for a long range of $\alpha$ and then contribute to $N_{eff}$.

On the other hand the red-shift when thermal $N_1$ becomes non-relativistic is given by:
\begin{equation}\label{r_redshift}
Z_{nr}^T + 1 = \frac{a_0}{a_{nr}}= \frac{a_0}{a_{f}} \frac{a_f}{a_{nr}}\,,
\end{equation}
where $a_f$ is the scale factor when $N_1$ freeze-out at the temperature $T_f$. The thermal $N_1$ momentum gets red-shifted
\begin{equation}\label{momentum_thermally}
    p_{N_1}^T\left( t \right) \ = \ T_{f} \frac{a_{f}}{a\left(t \right)}  \,.
\end{equation}
However we should note that as the thermal decoupling of $N_1$ 
happens
before the $N_2$ decay, the thermal $N_1$ will be cooler than the bath particles \cite{Patwardhan:2015kga}, and 
\begin{equation}\label{cooling_S1}
\frac{a_{f} }{a } \frac{T_{f}}{T} =\left( \frac{1}{\Delta}\frac{g_s(T)}{g_s(T_{f})}\right)^{1/3} \,,
\end{equation}
where $a$ and $T$ represent the scale factor and temperature for some epoch after $N_2$ decays. As non-thermal $N_1$ is produced during the injection of entropy due decay of $N_2$ we did not consider this contribution for non-thermal $N_1$ treatment.

As the thermal $N_1$ becomes non-relativistic when $p_{N_1}^T \sim m_{N_1}$, therefore from \autoref{momentum_thermally} and  \autoref{cooling_S1}, we can rewrite \autoref{r_redshift} as
\begin{equation}
    Z_{nr}^T + 1 =  \frac{m_{N_1}}{T_0} \left( \Delta  \frac{g_s \left(T_f \right)}{g_s \left(T_0 \right)} \right)^{1/3}\,.
\end{equation}

The parameter $\Delta$ plays an important role in $ Z_{nr}^T$. For $m_{N_1} = 2.0$ keV and for some $\alpha$ we only have a unique $\Delta$ that can reproduce $N_1$ as DM, such as for $\alpha = 0 $ we obtain numerically that we need of $\Delta \simeq 100$ to obtain $\Omega_{N_1} h^2 = 0.12$. As we discussed in \autoref{numerical} if we increase $\alpha$ we decrease the abundance of non-thermal $N_1$ which implies
that we need to decrease $\Delta$ in order to obtain $N_1$ as DM. Then we can obtain the lower value to $Z_{nr}^T$ for $m_{N_1} = 2.0$ keV if we take a scenario of the minimum allowed $\Delta$ value. This scenario can be reproduced if we assume that $N_1$ is only produced via freeze-out or if we assume that $\alpha$ is so large
that non-thermal contribution becomes irrelevant. In these cases we can assume that the lower value to $\Delta$ to provide the right amount of relic abundance for $N_1$ is $\Delta \simeq 19.0$, given by \autoref{Eq:DilutedRelic} if we take account the dilution factor $Y_{N_i} (T_0) = Y_{N_i} (T_f)/\Delta$
\begin{equation}
    \frac{\Omega_{N_1}^0 h^2}{0.12} \simeq \left(\frac{m_{N_1}}{2 \text{keV}}\right) \left(\frac{19.0}{\Delta}\right)\,,
\end{equation}
where we take $g_s(T_f) = 100$. For $\Delta = 19.0$ we obtain $Z_{NT}^T = 6.9 \times 10^7 \gg Z_{e}$, where $Z_{e} \simeq 3365$ is the red-shift at matter-radiation equality. This is an important result because it shows us that the $N_1$ produced via freeze-out and consequently diluted by the decay
of $N_2$ becomes non-relativistic before matter-radiation equality epoch.

\section{Evaluation of $\Delta N_{eff}$}\label{Ap:2}

Here we derive \autoref{effective_neutrino}. At temperature $T \lesssim 0.5$ MeV only the photon ($\gamma$), SM neutrinos ($\nu$) and non-thermal $N_1$ contribute to the radiation energy density of the universe. Then we can represent the energy density of radiation in that epoch as $\rho_R = \rho_\gamma + N^{SM}_\nu \rho_\nu + \rho_{N_1}$
The energy density for a ultra-relativistic particle (radiation) is given by:
\begin{equation}
    \rho_R = 
\left\{
\begin{alignedat}{2}
  g \frac{7}{8} \frac{ \pi^2 }{30} T^4&  \  \text{, Fermi-Dirac} \\
  g \frac{\pi^2}{30} T^4 &  \  \text{, Bose-Einstein}
\end{alignedat}
\right.
\end{equation}
where $g$ accounts for its spin degeneracy.

As we discussed in \autoref{sec:coupled}, we can write the energy density of non-thermal $N_1$ as function of the energy density of $N_2$ at the time that $N_1$ was produced (at $T_{RH}$), such that $\rho_{N_1} = f_{NT} \cdot \rho_{N_2} \left( T_{RH} \right)$, with $f_{NT}$ given by \autoref{fraction}. According to \cite{kolb_early_1990,PhysRevD.31.681}, between the epoch that $\rho_{N_2}$ starts to dominate the energy density of Universe until it decays (at $T_{RH}$) the radiation produced from decaying of $N_2$ is the dominant radiation component. Then we can assume that $\rho_{N_2} \left( T_{RH} \right) \simeq \rho_R \left( T_{RH} \right)$. As non-thermal $N_1$ does not thermalize anymore, its energy density only gets red-shifted. Then we can write the energy density of radiation as
\begin{equation}\label{radiation_density}
  \rho_R = \frac{\pi^2}{30} T^4 \left[ g_\gamma + \frac{7}{8} g_\nu  \left( \frac{T_\nu}{T}\right)^{4} \overbrace{\left[N^{SM}_\nu  + f_{NT} \cdot \frac{ g_e \left( T_{RH} \right)}{g_\nu} \frac{8}{7}\left( \frac{T}{T_\nu}\right)^{4}  \left( \frac{T_{N_1}^{RH}}{T}\right)^4\right]}^{N_{eff}}  \right]\,,
\end{equation}
where $T_\nu$ represents the temperature of SM neutrinos and $T_{N_1}^{RH}$ is the temperature of $N_1$ at the time of production (where $T_{N_1}^{RH} = T_{RH}$, after that $T_{N_1}^{RH}$ gets red-shifted). As $g_\nu = 2$, $T/T_\nu = (11/4)^{1/3}$ \cite{kolb_early_1990} and using the value of $f_{NT}$ given by \autoref{fraction} we can obtain $\Delta N_{eff}$ from \autoref{radiation_density} 
\begin{equation}
    \Delta N_{eff} = \frac{1}{3} \left( 1 + \alpha \right)^{-1} \cdot  g_e \left( T_{RH} \right) \frac{4}{7}\left( \frac{11}{4}\right)^{4/3}  \left( \frac{T_{N_1}^{RH}}{T}\right)^4\,.
\end{equation}

However we would like to know the $\Delta N_{eff}$ at matter-radiation equality epoch $T_e$. As the temperature of $N_1$ gets red-shifted after its production, we represent the temperature of $N_1$ at equality by $T_{N_1}^{e} = T_{N_1}^{RH} \times a_{RH}/a_e$. From entropy conservation we have $g_s(T_{RH}) T_{RH}^3 a_{RH} = g_s(T_e) T_e^3 a_e^3$, then
\begin{equation}
    \frac{1}{T_e^3} \left(\overbrace{ T_{N_1}^{RH} \frac{a_{RH}}{a_e} }^{T_{N_1}^e} \right)^3 = \frac{g_s \left( T_e \right)}{g_s \left( T_{RH} \right)}\,,
\end{equation}
and we obtain
\begin{equation}
    \Delta N_{eff}= \frac{4}{7} \left( \frac{11}{4} \right)^{4/3} \frac{1}{3} \left( 1 + \alpha 
    \right)^{-1} \left( \frac{g_{s, equality}^4}{g_{s, RH}} \right)^{1/3}\,.
\end{equation}
\bibliographystyle{JHEP}
\bibliography{biblio}

\providecommand{\href}[2]{#2}\begingroup\raggedright\begin{thebibliography}{10}

\bibitem{Hinshaw:2012aka}
{\scshape WMAP} collaboration, \emph{{Nine-Year Wilkinson Microwave Anisotropy
  Probe (WMAP) Observations: Cosmological Parameter Results}},
  \href{https://doi.org/10.1088/0067-0049/208/2/19}{\emph{Astrophys. J. Suppl.}
  {\bfseries 208} (2013) 19} [\href{https://arxiv.org/abs/1212.5226}{{\ttfamily
  1212.5226}}].

\bibitem{Aghanim:2018eyx}
{\scshape Planck} collaboration, \emph{{Planck 2018 results. VI. Cosmological
  parameters}},
  \href{https://doi.org/10.1051/0004-6361/201833910}{\emph{Astron. Astrophys.}
  {\bfseries 641} (2020) A6}
  [\href{https://arxiv.org/abs/1807.06209}{{\ttfamily 1807.06209}}].

\bibitem{Primack2001}
J.~R. Primack and M.~A.~K. Gross, \emph{Hot Dark Matter in Cosmology}. Springer
  Berlin Heidelberg, Berlin, Heidelberg, 2001,
  \href{https://doi.org/10.1007/978-3-662-04597-8\_12}{10.1007/978-3-662-04597-8\_12},
  [\href{https://arxiv.org/abs/astro-ph/0007165}{{\ttfamily
  astro-ph/0007165}}].

\bibitem{Bertone:2004pz}
G.~Bertone, D.~Hooper and J.~Silk, \emph{{Particle dark matter: Evidence,
  candidates and constraints}},
  \href{https://doi.org/10.1016/j.physrep.2004.08.031}{\emph{Phys. Rept.}
  {\bfseries 405} (2005) 279}
  [\href{https://arxiv.org/abs/hep-ph/0404175}{{\ttfamily hep-ph/0404175}}].

\bibitem{Arcadi:2017kky}
G.~Arcadi, M.~Dutra, P.~Ghosh, M.~Lindner, Y.~Mambrini, M.~Pierre et~al.,
  \emph{{The waning of the WIMP? A review of models, searches, and
  constraints}},
  \href{https://doi.org/10.1140/epjc/s10052-018-5662-y}{\emph{Eur. Phys. J. C}
  {\bfseries 78} (2018) 203}
  [\href{https://arxiv.org/abs/1703.07364}{{\ttfamily 1703.07364}}].

\bibitem{Bull:2015stt}
P.~Bull et~al., \emph{{Beyond $\Lambda$CDM: Problems, solutions, and the road
  ahead}}, \href{https://doi.org/10.1016/j.dark.2016.02.001}{\emph{Phys. Dark
  Univ.} {\bfseries 12} (2016) 56}
  [\href{https://arxiv.org/abs/1512.05356}{{\ttfamily 1512.05356}}].

\bibitem{Mohapatra:1974gc}
R.~N. Mohapatra and J.~C. Pati, \emph{{A Natural Left-Right Symmetry}},
  \href{https://doi.org/10.1103/PhysRevD.11.2558}{\emph{Phys. Rev. D}
  {\bfseries 11} (1975) 2558}.

\bibitem{Senjanovic:1975rk}
G.~Senjanovic and R.~N. Mohapatra, \emph{{Exact Left-Right Symmetry and
  Spontaneous Violation of Parity}},
  \href{https://doi.org/10.1103/PhysRevD.12.1502}{\emph{Phys. Rev. D}
  {\bfseries 12} (1975) 1502}.

\bibitem{Senjanovic:1978ev}
G.~Senjanovic, \emph{{Spontaneous Breakdown of Parity in a Class of Gauge
  Theories}}, \href{https://doi.org/10.1016/0550-3213(79)90604-7}{\emph{Nucl.
  Phys. B} {\bfseries 153} (1979) 334}.

\bibitem{Davidson:1978pm}
A.~Davidson, \emph{{$B-L$ as the fourth color within an $\mathrm{SU}(2)_L
  \times \mathrm{U}(1)_R \times \mathrm{U}(1)$ model}},
  \href{https://doi.org/10.1103/PhysRevD.20.776}{\emph{Phys. Rev. D} {\bfseries
  20} (1979) 776}.

\bibitem{Mohapatra:1980qe}
R.~N. Mohapatra and R.~E. Marshak, \emph{{Local B-L Symmetry of Electroweak
  Interactions, Majorana Neutrinos and Neutron Oscillations}},
  \href{https://doi.org/10.1103/PhysRevLett.44.1316}{\emph{Phys. Rev. Lett.}
  {\bfseries 44} (1980) 1316}.

\bibitem{Appelquist:2002mw}
T.~Appelquist, B.~A. Dobrescu and A.~R. Hopper, \emph{{Nonexotic Neutral Gauge
  Bosons}}, \href{https://doi.org/10.1103/PhysRevD.68.035012}{\emph{Phys. Rev.
  D} {\bfseries 68} (2003) 035012}
  [\href{https://arxiv.org/abs/hep-ph/0212073}{{\ttfamily hep-ph/0212073}}].

\bibitem{Basso:2008iv}
L.~Basso, A.~Belyaev, S.~Moretti and C.~H. Shepherd-Themistocleous,
  \emph{{Phenomenology of the minimal B-L extension of the Standard model: Z'
  and neutrinos}},
  \href{https://doi.org/10.1103/PhysRevD.80.055030}{\emph{Phys. Rev. D}
  {\bfseries 80} (2009) 055030}
  [\href{https://arxiv.org/abs/0812.4313}{{\ttfamily 0812.4313}}].

\bibitem{Khalil:2010iu}
S.~Khalil, \emph{{TeV-scale gauged B-L symmetry with inverse seesaw
  mechanism}}, \href{https://doi.org/10.1103/PhysRevD.82.077702}{\emph{Phys.
  Rev. D} {\bfseries 82} (2010) 077702}
  [\href{https://arxiv.org/abs/1004.0013}{{\ttfamily 1004.0013}}].

\bibitem{Singer:1980sw}
M.~Singer, J.~W.~F. Valle and J.~Schechter, \emph{{Canonical Neutral Current
  Predictions From the Weak Electromagnetic Gauge Group SU(3) X $u$(1)}},
  \href{https://doi.org/10.1103/PhysRevD.22.738}{\emph{Phys. Rev. D} {\bfseries
  22} (1980) 738}.

\bibitem{Montero:1992jk}
J.~C. Montero, F.~Pisano and V.~Pleitez, \emph{{Neutral currents and GIM
  mechanism in SU(3)-L x U(1)-N models for electroweak interactions}},
  \href{https://doi.org/10.1103/PhysRevD.47.2918}{\emph{Phys. Rev. D}
  {\bfseries 47} (1993) 2918}
  [\href{https://arxiv.org/abs/hep-ph/9212271}{{\ttfamily hep-ph/9212271}}].

\bibitem{Foot:1994ym}
R.~Foot, H.~N. Long and T.~A. Tran, \emph{{$SU(3)_L \otimes U(1)_N$ and
  $SU(4)_L \otimes U(1)_N$ gauge models with right-handed neutrinos}},
  \href{https://doi.org/10.1103/PhysRevD.50.R34}{\emph{Phys. Rev. D} {\bfseries
  50} (1994) R34} [\href{https://arxiv.org/abs/hep-ph/9402243}{{\ttfamily
  hep-ph/9402243}}].

\bibitem{Adhikari:2016bei}
M.~Drewes et~al., \emph{{A White Paper on keV Sterile Neutrino Dark Matter}},
  \href{https://doi.org/10.1088/1475-7516/2017/01/025}{\emph{JCAP} {\bfseries
  01} (2017) 025} [\href{https://arxiv.org/abs/1602.04816}{{\ttfamily
  1602.04816}}].

\bibitem{Kolb:1990vq}
E.~W. Kolb and M.~S. Turner, \emph{{The Early Universe}}, vol.~69. 1990.

\bibitem{Dror:2020jzy}
J.~A. Dror, D.~Dunsky, L.~J. Hall and K.~Harigaya, \emph{{Sterile Neutrino Dark
  Matter in Left-Right Theories}},
  \href{https://doi.org/10.1007/JHEP07(2020)168}{\emph{JHEP} {\bfseries 07}
  (2020) 168} [\href{https://arxiv.org/abs/2004.09511}{{\ttfamily
  2004.09511}}].

\bibitem{Bezrukov:2009th}
F.~Bezrukov, H.~Hettmansperger and M.~Lindner, \emph{{keV sterile neutrino Dark
  Matter in gauge extensions of the Standard Model}},
  \href{https://doi.org/10.1103/PhysRevD.81.085032}{\emph{Phys. Rev. D}
  {\bfseries 81} (2010) 085032}
  [\href{https://arxiv.org/abs/0912.4415}{{\ttfamily 0912.4415}}].

\bibitem{Nemevsek:2012cd}
M.~Nemevsek, G.~Senjanovic and Y.~Zhang, \emph{{Warm Dark Matter in Low Scale
  Left-Right Theory}},
  \href{https://doi.org/10.1088/1475-7516/2012/07/006}{\emph{JCAP} {\bfseries
  07} (2012) 006} [\href{https://arxiv.org/abs/1205.0844}{{\ttfamily
  1205.0844}}].

\bibitem{Foot:1992rh}
R.~Foot, O.~F. Hernandez, F.~Pisano and V.~Pleitez, \emph{{Lepton masses in an
  SU(3)-L x U(1)-N gauge model}},
  \href{https://doi.org/10.1103/PhysRevD.47.4158}{\emph{Phys. Rev. D}
  {\bfseries 47} (1993) 4158}
  [\href{https://arxiv.org/abs/hep-ph/9207264}{{\ttfamily hep-ph/9207264}}].

\bibitem{deSousaPires:1998jc}
C.~A. de~Sousa~Pires and O.~P. Ravinez, \emph{{Charge quantization in a chiral
  bilepton gauge model}},
  \href{https://doi.org/10.1103/PhysRevD.58.035008}{\emph{Phys. Rev. D}
  {\bfseries 58} (1998) 035008}
  [\href{https://arxiv.org/abs/hep-ph/9803409}{{\ttfamily hep-ph/9803409}}].

\bibitem{deSousaPires:1999ca}
C.~A. de~Sousa~Pires, \emph{{Remark on the vector - like nature of the
  electromagnetism and the electric charge quantization}},
  \href{https://doi.org/10.1103/PhysRevD.60.075013}{\emph{Phys. Rev. D}
  {\bfseries 60} (1999) 075013}
  [\href{https://arxiv.org/abs/hep-ph/9902406}{{\ttfamily hep-ph/9902406}}].

\bibitem{Dias:2005yh}
A.~G. Dias, C.~A. de~S.~Pires and P.~S. Rodrigues~da Silva, \emph{{Naturally
  light right-handed neutrinos in a 3-3-1 model}},
  \href{https://doi.org/10.1016/j.physletb.2005.09.028}{\emph{Phys. Lett. B}
  {\bfseries 628} (2005) 85}
  [\href{https://arxiv.org/abs/hep-ph/0508186}{{\ttfamily hep-ph/0508186}}].

\bibitem{Hoang:1995vq}
H.~N. Long, \emph{{The 331 model with right handed neutrinos}},
  \href{https://doi.org/10.1103/PhysRevD.53.437}{\emph{Phys. Rev. D} {\bfseries
  53} (1996) 437} [\href{https://arxiv.org/abs/hep-ph/9504274}{{\ttfamily
  hep-ph/9504274}}].

\bibitem{deS.Pires:2007gi}
C.~A. de~S.~Pires and P.~S. Rodrigues~da Silva, \emph{{Scalar Bilepton Dark
  Matter}}, \href{https://doi.org/10.1088/1475-7516/2007/12/012}{\emph{JCAP}
  {\bfseries 12} (2007) 012} [\href{https://arxiv.org/abs/0710.2104}{{\ttfamily
  0710.2104}}].

\bibitem{Mizukoshi:2010ky}
J.~K. Mizukoshi, C.~A. de~S.~Pires, F.~S. Queiroz and P.~S. Rodrigues~da Silva,
  \emph{{WIMPs in a 3-3-1 model with heavy Sterile neutrinos}},
  \href{https://doi.org/10.1103/PhysRevD.83.065024}{\emph{Phys. Rev. D}
  {\bfseries 83} (2011) 065024}
  [\href{https://arxiv.org/abs/1010.4097}{{\ttfamily 1010.4097}}].

\bibitem{Alvares:2012qv}
J.~D. Ruiz-Alvarez, C.~A. de~S.~Pires, F.~S. Queiroz, D.~Restrepo and P.~S.
  Rodrigues~da Silva, \emph{{On the Connection of Gamma-Rays, Dark Matter and
  Higgs Searches at LHC}},
  \href{https://doi.org/10.1103/PhysRevD.86.075011}{\emph{Phys. Rev. D}
  {\bfseries 86} (2012) 075011}
  [\href{https://arxiv.org/abs/1206.5779}{{\ttfamily 1206.5779}}].

\bibitem{Profumo:2013sca}
S.~Profumo and F.~S. Queiroz, \emph{{Constraining the $Z'$ mass in 331 models
  using direct dark matter detection}},
  \href{https://doi.org/10.1140/epjc/s10052-014-2960-x}{\emph{Eur. Phys. J. C}
  {\bfseries 74} (2014) 2960}
  [\href{https://arxiv.org/abs/1307.7802}{{\ttfamily 1307.7802}}].

\bibitem{Kelso:2013nwa}
C.~Kelso, C.~A. de~S.~Pires, S.~Profumo, F.~S. Queiroz and P.~S. Rodrigues~da
  Silva, \emph{{A 331 WIMPy Dark Radiation Model}},
  \href{https://doi.org/10.1140/epjc/s10052-014-2797-3}{\emph{Eur. Phys. J. C}
  {\bfseries 74} (2014) 2797}
  [\href{https://arxiv.org/abs/1308.6630}{{\ttfamily 1308.6630}}].

\bibitem{Dong:2014wsa}
P.~V. Dong, D.~T. Huong, F.~S. Queiroz and N.~T. Thuy, \emph{{Phenomenology of
  the 3-3-1-1 model}},
  \href{https://doi.org/10.1103/PhysRevD.90.075021}{\emph{Phys. Rev. D}
  {\bfseries 90} (2014) 075021}
  [\href{https://arxiv.org/abs/1405.2591}{{\ttfamily 1405.2591}}].

\bibitem{Cogollo:2014jia}
D.~Cogollo, A.~X. Gonzalez-Morales, F.~S. Queiroz and P.~R. Teles,
  \emph{{Excluding the Light Dark Matter Window of a 331 Model Using LHC and
  Direct Dark Matter Detection Data}},
  \href{https://doi.org/10.1088/1475-7516/2014/11/002}{\emph{JCAP} {\bfseries
  11} (2014) 002} [\href{https://arxiv.org/abs/1402.3271}{{\ttfamily
  1402.3271}}].

\bibitem{Kelso:2014qka}
C.~Kelso, H.~N. Long, R.~Martinez and F.~S. Queiroz, \emph{{Connection of
  $g-2_{\mu}$, electroweak, dark matter, and collider constraints on 331
  models}}, \href{https://doi.org/10.1103/PhysRevD.90.113011}{\emph{Phys. Rev.
  D} {\bfseries 90} (2014) 113011}
  [\href{https://arxiv.org/abs/1408.6203}{{\ttfamily 1408.6203}}].

\bibitem{Alves:2016fqe}
A.~Alves, G.~Arcadi, P.~V. Dong, L.~Duarte, F.~S. Queiroz and J.~W.~F. Valle,
  \emph{{Matter-parity as a residual gauge symmetry: Probing a theory of
  cosmological dark matter}},
  \href{https://doi.org/10.1016/j.physletb.2017.07.056}{\emph{Phys. Lett. B}
  {\bfseries 772} (2017) 825}
  [\href{https://arxiv.org/abs/1612.04383}{{\ttfamily 1612.04383}}].

\bibitem{Dong:2017zxo}
P.~V. Dong, D.~T. Huong, F.~S. Queiroz, J.~W.~F. Valle and C.~A.
  Vaquera-Araujo, \emph{{The Dark Side of Flipped Trinification}},
  \href{https://doi.org/10.1007/JHEP04(2018)143}{\emph{JHEP} {\bfseries 04}
  (2018) 143} [\href{https://arxiv.org/abs/1710.06951}{{\ttfamily
  1710.06951}}].

\bibitem{Ma:1998dn}
E.~Ma, \emph{{Pathways to naturally small neutrino masses}},
  \href{https://doi.org/10.1103/PhysRevLett.81.1171}{\emph{Phys. Rev. Lett.}
  {\bfseries 81} (1998) 1171}
  [\href{https://arxiv.org/abs/hep-ph/9805219}{{\ttfamily hep-ph/9805219}}].

\bibitem{Montero:2001ts}
J.~C. Montero, C.~A. De~S.~Pires and V.~Pleitez, \emph{{Neutrino masses through
  the seesaw mechanism in 3-3-1 models}},
  \href{https://doi.org/10.1103/PhysRevD.65.095001}{\emph{Phys. Rev. D}
  {\bfseries 65} (2002) 095001}
  [\href{https://arxiv.org/abs/hep-ph/0112246}{{\ttfamily hep-ph/0112246}}].

\bibitem{Cogollo:2009yi}
D.~Cogollo, H.~Diniz and C.~A. de~S.~Pires, \emph{{KeV right-handed neutrinos
  from type II seesaw mechanism in a 3-3-1 model}},
  \href{https://doi.org/10.1016/j.physletb.2009.05.060}{\emph{Phys. Lett. B}
  {\bfseries 677} (2009) 338}
  [\href{https://arxiv.org/abs/0903.0370}{{\ttfamily 0903.0370}}].

\bibitem{Dong:2008sw}
P.~V. Dong and H.~N. Long, \emph{{Neutrino masses and lepton flavor violation
  in the 3-3-1 model with right-handed neutrinos}},
  \href{https://doi.org/10.1103/PhysRevD.77.057302}{\emph{Phys. Rev. D}
  {\bfseries 77} (2008) 057302}
  [\href{https://arxiv.org/abs/0801.4196}{{\ttfamily 0801.4196}}].

\bibitem{Griest:1990kh}
K.~Griest and D.~Seckel, \emph{{Three exceptions in the calculation of relic
  abundances}}, \href{https://doi.org/10.1103/PhysRevD.43.3191}{\emph{Phys.
  Rev. D} {\bfseries 43} (1991) 3191}.

\bibitem{Hahn:2004fe}
T.~Hahn, \emph{{CUBA: A Library for multidimensional numerical integration}},
  \href{https://doi.org/10.1016/j.cpc.2005.01.010}{\emph{Comput. Phys. Commun.}
  {\bfseries 168} (2005) 78}
  [\href{https://arxiv.org/abs/hep-ph/0404043}{{\ttfamily hep-ph/0404043}}].

\bibitem{Seljak:2006qw}
U.~Seljak, A.~Makarov, P.~McDonald and H.~Trac, \emph{{Can sterile neutrinos be
  the dark matter?}},
  \href{https://doi.org/10.1103/PhysRevLett.97.191303}{\emph{Phys. Rev. Lett.}
  {\bfseries 97} (2006) 191303}
  [\href{https://arxiv.org/abs/astro-ph/0602430}{{\ttfamily
  astro-ph/0602430}}].

\bibitem{PhysRevD.31.681}
R.~J. Scherrer and M.~S. Turner, \emph{Decaying particles do not ``heat up''
  the universe}, \href{https://doi.org/10.1103/PhysRevD.31.681}{\emph{Phys.
  Rev. D} {\bfseries 31} (1985) 681}.

\bibitem{Cosme:2020mck}
C.~Cosme, M.~Dutra, T.~Ma, Y.~Wu and L.~Yang, \emph{{Neutrino Portal to FIMP
  Dark Matter with an Early Matter Era}},
  \href{https://arxiv.org/abs/2003.01723}{{\ttfamily 2003.01723}}.

\bibitem{Hasegawa:2019jsa}
T.~Hasegawa, N.~Hiroshima, K.~Kohri, R.~S.~L. Hansen, T.~Tram and S.~Hannestad,
  \emph{{MeV-scale reheating temperature and thermalization of oscillating
  neutrinos by radiative and hadronic decays of massive particles}},
  \href{https://doi.org/10.1088/1475-7516/2019/12/012}{\emph{JCAP} {\bfseries
  12} (2019) 012} [\href{https://arxiv.org/abs/1908.10189}{{\ttfamily
  1908.10189}}].

\bibitem{Patwardhan:2015kga}
A.~V. Patwardhan, G.~M. Fuller, C.~T. Kishimoto and A.~Kusenko, \emph{{Diluted
  equilibrium sterile neutrino dark matter}},
  \href{https://doi.org/10.1103/PhysRevD.92.103509}{\emph{Phys. Rev. D}
  {\bfseries 92} (2015) 103509}
  [\href{https://arxiv.org/abs/1507.01977}{{\ttfamily 1507.01977}}].

\bibitem{Fuller:2011qy}
G.~M. Fuller, C.~T. Kishimoto and A.~Kusenko, \emph{{Heavy sterile neutrinos,
  entropy and relativistic energy production, and the relic neutrino
  background}},  \href{https://arxiv.org/abs/1110.6479}{{\ttfamily 1110.6479}}.

\bibitem{Cao:2016uur}
Q.-H. Cao and D.-M. Zhang, \emph{{Collider Phenomenology of the 3-3-1 Model}},
  \href{https://arxiv.org/abs/1611.09337}{{\ttfamily 1611.09337}}.

\bibitem{Coutinho:2013lta}
Y.~A. Coutinho, V.~Salustino Guimar\~aes and A.~A. Nepomuceno, \emph{{Bounds on
  Z' from 3-3-1 model at the LHC energies}},
  \href{https://doi.org/10.1103/PhysRevD.87.115014}{\emph{Phys. Rev. D}
  {\bfseries 87} (2013) 115014}
  [\href{https://arxiv.org/abs/1304.7907}{{\ttfamily 1304.7907}}].

\bibitem{Cogollo:2020afo}
D.~Cogollo, F.~F. Freitas, C.~A. de~S.~Pires, Y.~M. Oviedo-Torres and
  P.~Vasconcelos, \emph{{Deep learning analysis of the inverse seesaw in a
  3-3-1 model at the LHC}},
  \href{https://doi.org/10.1016/j.physletb.2020.135931}{\emph{Phys. Lett. B}
  {\bfseries 811} (2020) 135931}
  [\href{https://arxiv.org/abs/2008.03409}{{\ttfamily 2008.03409}}].

\bibitem{deMelo:2021ers}
T.~B. de~Melo, S.~Kovalenko, F.~S. Queiroz, C.~Siqueira and Y.~S. Villamizar,
  \emph{{Rare Kaon Decay to Missing Energy: Implications of the NA62 Result for
  a $Z^\prime$ Model}},  \href{https://arxiv.org/abs/2102.06262}{{\ttfamily
  2102.06262}}.

\bibitem{Froustey:2020mcq}
J.~Froustey, C.~Pitrou and M.~C. Volpe, \emph{{Neutrino decoupling including
  flavour oscillations and primordial nucleosynthesis}},
  \href{https://doi.org/10.1088/1475-7516/2020/12/015}{\emph{JCAP} {\bfseries
  12} (2020) 015} [\href{https://arxiv.org/abs/2008.01074}{{\ttfamily
  2008.01074}}].

\bibitem{Abazajian:2019eic}
K.~Abazajian et~al., \emph{{CMB-S4 Science Case, Reference Design, and Project
  Plan}},  \href{https://arxiv.org/abs/1907.04473}{{\ttfamily 1907.04473}}.

\bibitem{Merle:2013wta}
A.~Merle, V.~Niro and D.~Schmidt, \emph{{New Production Mechanism for keV
  Sterile Neutrino Dark Matter by Decays of Frozen-In Scalars}},
  \href{https://doi.org/10.1088/1475-7516/2014/03/028}{\emph{JCAP} {\bfseries
  03} (2014) 028} [\href{https://arxiv.org/abs/1306.3996}{{\ttfamily
  1306.3996}}].

\bibitem{kolb_early_1990}
E.~W. Kolb and M.~S. Turner, \emph{The Early Universe}, vol.~69. Westview
  Press.

\bibitem{Borgani:1996ag}
S.~Borgani, A.~Masiero and M.~Yamaguchi, \emph{{Light gravitinos as mixed dark
  matter}}, \href{https://doi.org/10.1016/0370-2693(96)00956-2}{\emph{Phys.
  Lett. B} {\bfseries 386} (1996) 189}
  [\href{https://arxiv.org/abs/hep-ph/9605222}{{\ttfamily hep-ph/9605222}}].

\bibitem{Lin:2000qq}
W.~B. Lin, D.~H. Huang, X.~Zhang and R.~H. Brandenberger, \emph{{Nonthermal
  production of WIMPs and the subgalactic structure of the universe}},
  \href{https://doi.org/10.1103/PhysRevLett.86.954}{\emph{Phys. Rev. Lett.}
  {\bfseries 86} (2001) 954}
  [\href{https://arxiv.org/abs/astro-ph/0009003}{{\ttfamily
  astro-ph/0009003}}].

\bibitem{10.1111/j.1365-2966.2011.20200.x}
M.~R. Lovell, V.~Eke, C.~S. Frenk, L.~Gao, A.~Jenkins, T.~Theuns et~al.,
  \emph{{The haloes of bright satellite galaxies in a warm dark matter
  universe}},
  \href{https://doi.org/10.1111/j.1365-2966.2011.20200.x}{\emph{Monthly Notices
  of the Royal Astronomical Society} {\bfseries 420} (2012) 2318}
  [\href{https://arxiv.org/abs/https://academic.oup.com/mnras/article-pdf/420/3/2318/3020178/mnras0420-2318.pdf}{{\ttfamily
  https://academic.oup.com/mnras/article-pdf/420/3/2318/3020178/mnras0420-2318.pdf}}].

\bibitem{Bowman:2018yin}
J.~D. Bowman, A.~E.~E. Rogers, R.~A. Monsalve, T.~J. Mozdzen and N.~Mahesh,
  \emph{{An absorption profile centred at 78 megahertz in the sky-averaged
  spectrum}}, \href{https://doi.org/10.1038/nature25792}{\emph{Nature}
  {\bfseries 555} (2018) 67}
  [\href{https://arxiv.org/abs/1810.05912}{{\ttfamily 1810.05912}}].

\bibitem{Chatterjee:2019jts}
A.~Chatterjee, P.~Dayal, T.~R. Choudhury and A.~Hutter, \emph{{Ruling out 3 keV
  warm dark matter using 21 cm EDGES data}},
  \href{https://doi.org/10.1093/mnras/stz1444}{\emph{Mon. Not. Roy. Astron.
  Soc.} {\bfseries 487} (2019) 3560}
  [\href{https://arxiv.org/abs/1902.09562}{{\ttfamily 1902.09562}}].

\bibitem{Dayal:2015vca}
P.~Dayal, T.~R. Choudhury, V.~Bromm and F.~Pacucci, \emph{{Reionization and
  Galaxy Formation in Warm Dark Matter Cosmologies}},
  \href{https://doi.org/10.3847/1538-4357/836/1/16}{\emph{Astrophys. J.}
  {\bfseries 836} (2017) 16}
  [\href{https://arxiv.org/abs/1501.02823}{{\ttfamily 1501.02823}}].

\bibitem{Schneider:2018xba}
A.~Schneider, \emph{{Constraining noncold dark matter models with the global
  21-cm signal}}, \href{https://doi.org/10.1103/PhysRevD.98.063021}{\emph{Phys.
  Rev. D} {\bfseries 98} (2018) 063021}
  [\href{https://arxiv.org/abs/1805.00021}{{\ttfamily 1805.00021}}].

\bibitem{Hasenkamp:2012ii}
J.~Hasenkamp and J.~Kersten, \emph{{Dark radiation from particle decay:
  cosmological constraints and opportunities}},
  \href{https://doi.org/10.1088/1475-7516/2013/08/024}{\emph{JCAP} {\bfseries
  08} (2013) 024} [\href{https://arxiv.org/abs/1212.4160}{{\ttfamily
  1212.4160}}].

\end{thebibliography}\endgroup
\end{document}